\documentclass[prb,showpacs,preprintnumbers,amsfonts,amssymb,amsmath,floats,twocolumn,aps]{revtex4}
 
\usepackage[pdftex]{graphicx}
\usepackage[pdftex,colorlinks=true,linkcolor=blue,citecolor=blue,filecolor=blue,baseurl=empty]{hyperref}
\usepackage{dcolumn}
\usepackage{bm}
\usepackage{color}
\usepackage{amsmath}

\newcommand{\ff}[1]{{\boldsymbol #1}}
\newcommand{\ca}[1]{{\cal #1}}

\newcommand{\bi}{\begin{itemize}}
\newcommand{\ei}{\end{itemize}}
\newcommand{\be}{\begin{equation}}
\newcommand{\ee}{\end{equation}}
\newcommand{\lr}[1]{\left(#1\right)}
\newcommand{\matrixel}[3]{\big< #1 \vphantom{#2#3} \big| #2 \big| #3 \vphantom{#1#2} \big>} 
\newcommand{\ket}[1]{\big| #1 \big>} 
\newcommand{\bra}[1]{\big< #1 \big|} 
\newcommand{\avg}[1]{\big< #1 \big>} 
\newcommand{\braket}[2]{\big<#1 \vphantom{#2} \big| #2 \vphantom{#1} \big>} 

\newcommand{\norm}[1]{\left|#1\right|^2}
\newcommand{\rec}[1]{\frac{1}{#1}}

\begin{document}

\title{Multiplons in the two-hole excitation spectra of the one-dimensional Hubbard model}

\author{Roman Rausch and Michael Potthoff}
\affiliation{I. Institute for Theoretical Physics, University of Hamburg, Jungiusstra\ss{}e 9, 20355 Hamburg, Germany}

\begin{abstract}
Using the density-matrix renormalization group in combination with the Chebyshev polynomial expansion technique, we study the two-hole excitation spectrum of the one-dimensional Hubbard model in the entire filling range from the completely occupied band ($n=2$) down to half-filling ($n=1$). 
For strong interactions, the spectra reveal multiplon physics, i.e., relevant final states are characterized by two (doublon), three (triplon), four (quadruplon) and more holes, potentially forming stable compound objects or resonances with finite lifetime. 
These give rise to several satellites in the spectra with largely different spectral weights as well as to different two-hole, doublon-hole, two-doublon etc.\ continua. 
The complex multiplon phenomenology is analyzed by interpreting not only local and $k$-resolved two-hole spectra but also three- and four-hole spectra for the Hubbard model and by referring to effective low-energy models. 
In addition, a filter-operator technique is presented and applied which allows to extract specific information on the final states at a given excitation energy.
While multiplons composed of an odd number of holes do neither form stable compounds nor well-defined resonances unless a nearest-neighbor density interaction $V$ is added to the Hamiltonian, the doublon and the quadruplon are well-defined resonances. 
The $k$-resolved four-hole spectrum at $n=2$ represents an interesting special case where a completely stable quadruplon turns into a resonance by merging with the doublon-doublon continuum at a critical wave vector.
For all fillings with $n > 1$, the doublon lifetime is strongly $k$-dependent and is even infinite at the Brillouin zone edges as demonstrated by $k$-resolved two-hole spectra. 
This can be traced back to the ``hidden'' charge-SU(2) symmetry of the model which is explicitly broken off half-filling and gives rise to a massive collective excitation, even for arbitrary higher-dimensional but bipartite lattices. 
\end{abstract}
 


\maketitle

\section{Introduction}
\label{sec:intro}

The prime example for emergent phenomena in systems of elementary interacting particles is the formation of compound objects with new properties.
Compound objects can be stable, i.e., described by exact eigenstates of the microscopic Hamiltonian.
Common examples are the hydrogen atom, stable molecules, or mesons and baryons.
In other cases, they are described by bound states interacting with a continuum. 
Those excitations acquire a finite lifetime and show up as resonances of finite width in various spectra, \cite{Maj31}
such as Fano resonances \cite{Fan35} in the photo-absorption spectrum caused by autoionizing states of two-electron atoms, or the Auger effect in atoms and solids with inner-shell holes. \cite{Mei22,Aug23}

Recently, considerable attention has been devoted to repulsively bound pairs of particles \cite{WTL+06} formed in a Bose-Hubbard model. \cite{Blo05,JZ05}
If the on-site Hubbard interaction $U$ is sufficiently large compared to the width of the free Bloch band $W$, two bosons prepared on the same site move as a bound pair through the lattice. 
In an otherwise empty conduction band, this doublon is a completely stable compound object despite the fact that the Hubbard $U$ is repulsive. 
This is a consequence of energy conservation which implies that there is no phase space available for a dissociation into two particles moving independently through the lattice. \cite{HDD06}
The latter situation is described by states which belong to a scattering continuum at energies within twice the bandwidth $2W$.

If there is, on the other hand, a finite concentration of bosons or fermions \cite{JSG+08} in the conduction band, the doublon may decay by transferring its energy $\sim U$ in a high-order scattering process to several low-lying excitations of the continuum. 
This implies a finite lifetime of the doublon and poses an intricate many-body problem.
The lifetime has been measured \cite{SGJ+10} for a well-controlled Fermi gas trapped in an optical lattice and was estimated theoretically for Fermi \cite{SGJ+10,LP13} and for Bose systems. \cite{CGK12}
Studies of the real-time dynamics of an initially doubly occupied state
$| \Psi(t) \rangle = c^{\dagger}_{j\uparrow}(t) c^{\dagger}_{j\downarrow}(t) | 0 \rangle$,
where $|0\rangle$ is the vacuum state or the few-particle ground state, have been done with exact-diagonalization methods. \cite{PSAF07,HP12} 
For $|0\rangle$ being the ground state of the extended Hubbard model at half-filling, time-dependent density-matrix renormalization group (DMRG) \cite{WF04,DKSV04} has been used to study the decay of a single doublon-holon pair \cite{ARF+08,DAD12} and of a large number of doublons. \cite{ES12}

In the context of condensed-matter systems, the idea of repulsively bound pairs goes back to the so-called Cini-Sawatzky theory developed in the late 1970s. \cite{Cin77,Saw77}
This addresses the energy-dependent cross section for core-valence-valence (CVV) Auger decay for a system with an initial core hole and a fully occupied conduction band described by a strongly correlated (Fermi-)Hubbard model $H$.
Adopting a number of standard approximations (see Refs.\ \onlinecite{Pot01b,PWN+00} for a discussion), such as the sudden approximation and constant transition-matrix elements, and disregarding core-hole screening effects, the bare Auger spectrum $I_{\rm AES}(\omega) \propto (- 1/\pi) \, \text{Im} \, G_{\rm 2-hole}(\omega + i 0^{+})$ is given in terms of the local two-hole Green's function:
\be
G_{\rm 2-hole}(\omega) 
=
\bra{0} c^{\dagger}_{i\downarrow} c^{\dagger}_{i\uparrow} \frac{1}{\omega - (E_{0} - H)}  c_{i\uparrow} c_{i\downarrow} \ket{0}
\ee
Here, $E_{0}$ is the ground-state energy, $\omega$ is the two-hole excitation energy, $i$ is the lattice site at which the two final-state holes are created, and $|0\rangle$ is the hole vacuum, i.e. the fully occupied band.
Note that due to the core hole involved in the transition, the Auger process can be assumed as local to a good approximation.

With a particle-hole transformation, the immediate relation to doublon dynamics becomes obvious:
The Auger spectrum consists of a continuum of scattering states (``band-like part'') related to the independent motion of the two final-state holes and, for sufficiently strong Coulomb repulsion $U$, of the so-called Cini-Sawatzky satellite at higher two-hole binding energies, which corresponds to repulsively bound two-hole states. 
It has a finite dispersion but zero width, so that the compound object of two holes moves through the lattice, but cannot decay.
Calculations can be done analytically for the simple two-particle scattering problem. \cite{Cin77,Saw77}

However, beyond the case of a fully occupied conduction band, there is not much known about the two-hole excitation spectrum:
The Cini-Sawatzky theory can be generated by a ladder approximation within diagrammatic perturbation theory. 
This idea has been used \cite{Nol90} to extend the theory to partially filled conduction bands, albeit in an essentially {\em ad hoc} way. 
Some progress was made by summing an extended diagram class which dominates in the low-hole-density limit \cite{Cin79,DK84} as well as by consideration of diagrams up second order in weak-coupling perturbation theory. \cite{GDDS81} 
Except for straightforward exact-diagonalization studies, \cite{OTSJ86,CV86} which are plagued by strong finite-size artifacts, however, there has not been any systematic study of the two-hole excitation spectrum in the interesting regime of finite hole densities up to the half-filled conduction band where a strong Hubbard interaction is expected to be most effective.

With the present paper we discuss the two-hole spectrum in the entire filling range for the one-dimensional Hubbard model (including its extended variant) in the strong-coupling regime.
To this end, we employ a matrix-product state (MPS) implementation of DMRG \cite{Sch11} in combination with the Chebyshev polynomial expansion technique. \cite{WWAF06,SR94,SR97} 
This method \cite{HWMcC+11} provides us with essentially exact two-hole excitation spectra for lattices with several tens of sites.
We also propose a filter-operator technique which can be implemented straightforwardly within the framework of the Chebyshev-DMRG technique and which is shown to be a very helpful tool to characterize the final states at a given excitation energy. 

Our central goal is to study the fate of the concept of a repulsively bound pair of holes if it is allowed to interact with the states of the continuum of particle-hole excitations and how this manifests itself in the two-hole spectrum. 
This makes contact with several questions, for example:
Is the doublon a well-defined resonance or even a stable compound object for a partially occupied band or does it decay immediately?
How does its excitation energy, its dispersion and its weight change with the filling and can the excitation be traced continuously down to half-filling?
Studying to those questions will put the clear physical picture of the Cini-Sawatzky model to a test.

In an effort to answer those questions, one soon recognizes that the two-hole spectrum is actually more complex than expected and exhibits several satellite features and continua.
Those arise from heavier objects consisting of three and more holes. 
The question for the stability of these heavier multiplons arises immediately. 
To study the triplon and the quadruplon, in particular, it is helpful to consider higher-order spectroscopies as well, such as three- and four-hole spectroscopy for the Hubbard model or a two-hole spectroscopy for the hard-core boson model that captures the relevant low-energy physics close to complete filling.

The paper is organized as follows:
The next section \ref{sec:mod} briefly introduces the model and discusses Chebyshev-DMRG method.
Results are presented and discussed in Sec.\ \ref{sec:res}. 
We start by recalling the physics of the completely filled band in Sec.\ \ref{sec:filled}. 
For the partially filled band, we first give an overview in Sec.\ \ref{sec:part} before we discuss the main spectral features separately, namely the doublon (Sec.\ \ref{sec:doublon}), the band-like part or two-hole continuum (Sec.\ \ref{sec:band}), and the triplon and the quadruplon in Secs.\ \ref{sec:triplon} and \ref{sec:quad}. 
In Sec.\ \ref{sec:filter} we introduce the new analysis tool based on filter operators and return to the question of the doublon lifetime in Sec.\ \ref{sec:res:kdependence}. 
Finally, a general and concluding discussion is given in Sec.\ \ref{sec:con}.

\section{Model and method}
\label{sec:mod}

Using standard notations, the Hubbard model reads \cite{Gia04}
\be
H 
= 
- t \sum_{\langle ij \rangle , \sigma} \lr{c^{\dagger}_{i\sigma} c_{j\sigma} + \text{H.c.}} 
+ 
U \sum_{i} n_{i\uparrow} n_{i\downarrow}
\; .
\label{eq:ham}
\ee
We consider the model on a one-dimensional lattice of $L$ sites with open boundary conditions and $N$ electrons with $L \le N \le 2L$, i.e., for fillings $n=N/L$ ranging from half-filling $n=1$ to the limit of the completely occupied band $n=2$. 
In Eq.\ (\ref{eq:ham}), 
$c^{\dagger}_{i\sigma}$ creates an electron at lattice site $i$ with spin projection $\sigma = \uparrow,\downarrow$, and $n_{i\sigma}=c^{\dagger}_{i\sigma}c_{i\sigma}$ is the occupation-number operator. 
Summation over ordered pairs of nearest neighbors is denoted by $\langle ij \rangle$. 
Furthermore, $U \ge 0$ is the strength of the on-site Coulomb repulsion. 
Energy units are fixed by setting the nearest-neighbor hopping $t=1$, and the resulting width of the free tight-binding band is $W=4$.
It turns out as instructive to study the extended Hubbard model in addition.
The Hamiltonian,
\be
  H_{\text{ext}} = H + V \sum_{\left<ij\right>,\sigma,\sigma'} n_{i\sigma} n_{j\sigma'}
  \;  ,
\label{eq:hamext}  
\ee
includes an additional nearest-neighbor Coulomb interaction of strength $V$.
DMRG has been previously employed to study its ground-state phase diagram. \cite{Jec02,EN08}
One-particle spectral properties \cite{BGJ04,FH09,Koh10} 
as well as
dynamical spin- and charge-correlation functions \cite{BJ07,NSO08} have been computed using the dynamical DMRG method
\cite{Jec02} and time-dependent DMRG. \cite{WF04}

Our goal is to compute the local two-particle spectral function
\be
A_2\lr{\omega} = A_{\rm 2-hole}\lr{\omega} + A_{\rm 2-particle}\lr{\omega}
\: ,
\ee
which consists of the two-hole excitation spectrum
\begin{widetext}
\be
A_{\rm 2-hole}\lr{\omega} 
= 
\sum_n \norm{\matrixel{n,N-2}{c_{i\uparrow}c_{i\downarrow}}{0,N}} 
\delta\lr{\omega + 2 \mu^{\lr{N-1}} -\lr{E_0^{\lr{N}}-E_n^{\lr{N-2}}}}
\: ,
\label{eq:A2hole}
\ee
with excitation energies $\omega \le 0$, related to Auger-electron spectroscopy, 
and of the two-particle excitation spectrum
\be
A_{\rm 2-particle}\lr{\omega} = \sum_n \norm{\matrixel{n,N+2}{c^{\dagger}_{i\downarrow}c^{\dagger}_{i\uparrow}}{0,N}} 
\delta\lr{\omega + 2 \mu^{\lr{N+1}} -\lr{E_n^{\lr{N+2}}-E_0^{\lr{N}}}}
\: ,
\label{eq:A2particle}
\ee
\end{widetext}
with excitation energies $\omega \ge 0$, related to appearance-potential spectroscopy. \cite{PWN+00}
Here, $| 0, N \rangle$ is the ground state of $H$ in the $N$-particle invariant subspace with the corresponding ground-state energy $E_{0}^{(N)}$, while $| n, N \pm 2\rangle$ denotes the $n$-th excited state with energy $E_{n}^{(N\pm 2)}$ in the subspace two additional holes or particles, respectively.
The corresponding chemical potentials $\mu^{(N\pm1)}$ are fixed by 
\begin{eqnarray}
\mu^{\lr{N-1}} &=& (E_0^{\lr{N}} - E_0^{\lr{N-2}} ) / 2
\: , 
\nonumber \\
\mu^{\lr{N+1}} &=& (E_0^{\lr{N+2}} - E_0^{\lr{N}}) / 2
\: .
\label{eq:mu}
\end{eqnarray}
Particle-hole symmetry implies that the two-hole spectrum at filling $n$ is related to the two-particle spectrum at filling $2-n$ via
\be
  A_{\text{2-hole}}\lr{\omega,n} = A_{\text{2-particle}}\lr{-\omega,2-n} \: .
\label{eq:phsym}  
\ee
It is therefore sufficient to study the two-hole and the two-particle spectrum for fillings at and above half-filling: $1\leq n \leq 2$.

We employ DMRG in combination with the Chebyshev expansion technique. \cite{WWAF06,SR94,SR97}
The main idea is to expand the spectral function into Chebyshev polynomials $T_k(x)$ which are defined on the interval $0\le x \le 1$, either via a recursion relation,
\be
  T_{k}\lr{x} = 2x T_{k-1}\lr{x} - T_{k-2}\lr{x}
\label{Cheb_recursion}
\ee
with $T_0(x) \equiv1$, $T_1(x) \equiv x$, or with the weight function $w\lr{x}=1/\pi\sqrt{1-x^2}$ from the orthogonality relation
\be
\int_{-1}^{1} dx~ \rec{\pi\sqrt{1-x^2}} T_k\lr{x} T_l\lr{x} = \rec{2}\lr{1+\delta_{k0}} \delta_{kl}
\: .
\label{Cheb_ortho}
\ee
An application requires a proper rescaling of the spectrum of the Hamiltonian $H$ to ensure that its rescaled eigenvalues have modulus less then unity. 
This can be most simply achieved with the linear transformation $\widetilde{H} = \lr{H-b} / a$ where $a = \lr{E_{\text{max}}-E_{\text{min}}}/2$ and $b = \lr{E_{\text{max}} + E_{\text{min}}}/2$ and adding a small safety padding to avoid touching the edges.
The extremal eigenvalues $E_{\text{min/max}}$ of $H$ are obtained from a standard DMRG ground-state calculation in the $N\pm 2$ subspace.

A spectral function $A(x) = \matrixel{0}{B_{2} \delta (x-\widetilde{H}) B_{1}}{0}$ of the rescaled frequency $x$, defined in terms of operators $B_{1}$ and $B_{2}$, is obtained as
\be
A(x) 
= 
\rec{\pi\sqrt{1-x^2}} \lr{\mu_0 + 2\sum_{k=1}^{\infty} \mu_k T_k\lr{x}}
\: .
\label{eq:acheb}
\ee
Using Eq.\ (\ref{Cheb_ortho}), the moments $\mu_{k}$ can be expressed as
\be
\mu_k 
= 
\matrixel{0}{B_{2} ~T_k (\widetilde{H})~ B_{1}}{0}
\ee
and are calculated as $\mu_{k} = \langle 0 | B_{2} | t_{k} \rangle$ from states $|t_{k} \rangle$ which, by making use of the recursion relation Eq.\ (\ref{Cheb_recursion}), can be constructed recursively, 
\be
\ket{t_k} 
= 
2 \widetilde{H} \ket{t_{k-1}} - \ket{t_{k-2}} 
\; ,
\label{Cheb_recursion_vectors}
\ee
starting from $\ket{t_0} = B_{1} \ket{0}$ and $\ket{t_1} = \widetilde{H} \ket{t_0}$.
A final back-scaling yields the desired $A(\omega)$.

Naive truncation of the series in Eq.\ (\ref{eq:acheb}) at $k=M<\infty$ causes so-called Gibbs oscillations close to sharp features in the spectral function. \cite{WWAF06}
Those can be attenuated by several orders of magnitude when multiplying the moments $\mu_k$ by correction factors $g^{\lr{M}}_k$. 
This amounts to a convolution of $A(x)$ with a kernel and results in the kernel-polynomial approximation
\be
A(x) \approx \rec{\pi\sqrt{1-x^2}} \lr{\mu_0g^{\lr{M}}_0 + 2\sum_{k=1}^{M-1} \mu_kg^{\lr{M}}_k T_k\lr{x}}
\: .
\ee
The two main choices for $g^{\lr{M}}_k$ are the Lorentz and the Jackson kernel \cite{WWAF06} corresponding to Lorentzian or Gaussian broadening of the $\delta$-peaks, respectively.
While the Lorentz kernel preserves the analytical properties of the spectral function, the Jackson kernel offers a somewhat better resolution. 
When working with matrix-product states, the effort in calculating the moments becomes quite significant, and we therefore opt for the latter throughout this work in order to obtain more spectral information for a given value of $M$.
We have not seen any need to apply linear prediction\cite{Sch11,GTVHE14,WCPS14} in the present study.

We apply the Chebyshev expansion method to exact many-body states for moderately large Hilbert spaces and, 
in case of large systems, combine it with the concept of matrix-product states \cite{Sch11} as suggested in Ref.\ \onlinecite{HWMcC+11}.
An MPS is obtained by a matrix-product ansatz for the coefficients of the many-body wave function in the product basis of states $\{\ket{s_{i}}\}$ spanning the local Hilbert space at a site $i$: 
\be
\ket{\Psi} = \sum_{s_1,s_2,\ldots,s_L} \underline{A}^{(s_1)}\underline{A}^{(s_2)} \ldots \underline{A}^{(s_L)} \ket{s_1}\ket{s_2} \ldots \ket{s_L}.
\ee
For the Hubbard model, $\ket{s_i}=\ket{0},\ket{\uparrow},\ket{\downarrow}\ket{\uparrow\downarrow}$. 
This ansatz allows an efficient truncation of the wave function by truncating the individual matrices. 
Furthermore, many operations can be applied locally with subsequent sweeps through the chain until convergence is reached.
However, the MPS representation also comes at a cost, as the application of $\widetilde{H}$ and the sum of two states in Eq.\ (\ref{Cheb_recursion_vectors}) can usually be done in an approximate way only.
Namely, to find an optimal representation of the left-hand side of Eq.\ (\ref{Cheb_recursion_vectors}), we perform a variational compression, \cite{Sch11} sweeping along the chain and optimizing one site at a time until $\|\ket{t_{k}}-(2\widetilde{H} \ket{t_{k-1}}-\ket{t_{k-2}})\|^2<\epsilon$ with a given tolerance $\epsilon$.
Throughout this work $\epsilon=10^{-4}$ is used. 
Note that it is possible to set up two environments for the overlaps $\matrixel{t_k}{2\widetilde{H}}{t_{k-1}}$ and $\braket{t_k}{t_{k-2}}$, and impose the above condition locally without the need of carrying out the subtractions one at a time.
As an initial guess to start the sweep we use the Chebyshev vector from the previous iteration. 
If convergence is not reached after four half-sweeps, we switch to a two-site algorithm which allows both to escape a local minimum and to dynamically increase the matrix dimensions, while keeping track of conserved quantum numbers.

The application of a local operator $B_{1}=c_{i\uparrow}c_{i\downarrow}$ to the ground state $\ket{0}$ creates a local perturbation which propagates through the lattice when repeatedly applying the Hamiltonian. 
The entanglement entropy therefore grows with the iteration index $k$, i.e., the state $\ket{t_k}$ in general requires a higher bond dimension than $\ket{t_{k-1}}$ for a correct representation of the wave function. 
We typically observe a logarithmic increase of the entanglement entropy with $k$, consistent with the findings of Ref.\ \onlinecite{CC07}.
The maximal iteration depth is hence limited in DMRG as, assuming a fixed error, one eventually either runs out of memory or the procedure slows down beyond practicability.

The Chebyshev method has an approximately uniform energy resolution $\delta E$ given by $\delta E \sim \Delta E/M$ where $\Delta E = E_{\text{max}}-E_{\text{min}}$ is the total bandwidth. \cite{WWAF06}
In order to be able to compare the spectra for various fillings and parameters, we therefore fix $\delta E$ as a measure of resolution rather than $M$ in all calculations.

\section{Results}
\label{sec:res}

\subsection{Completely filled band}
\label{sec:filled}

\begin{figure}[t]
\includegraphics[width=\columnwidth]{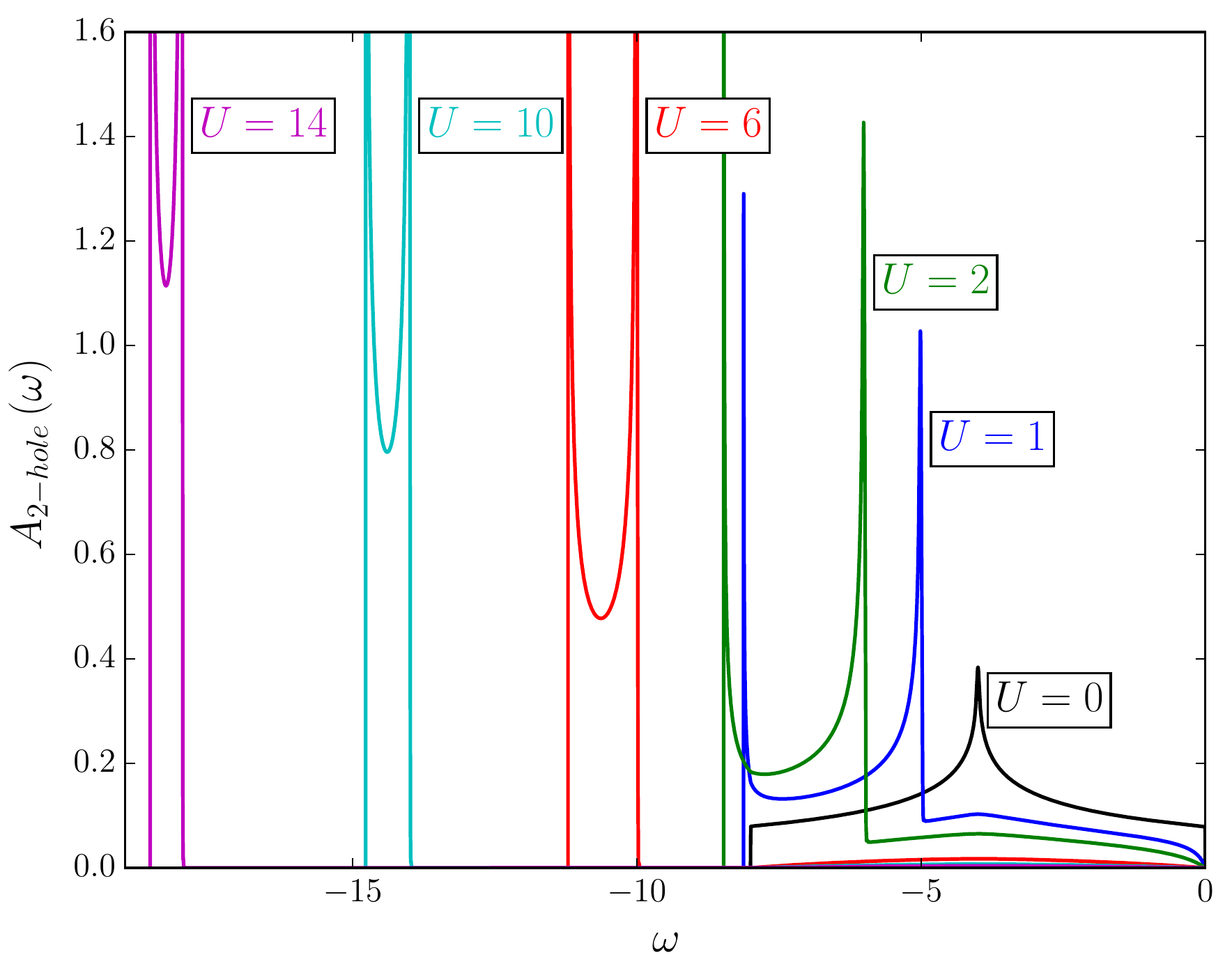}
\caption{
Two-hole excitation spectrum of the one-dimensional Hubbard model for a completely filled band (filling $n=2$) and several $U$ as indicated.
Energy units are fixed by the nearest-neighbor hopping $t=1$. 
Results as obtained by the Chebyshev expansion method using exact states for $L=1000$ lattice sites and periodic boundary conditions. 
Energy resolution: $\delta E = 0.01t$, obtained with $M=799$ to $M=1856$ Chebyshev moments.
}
\label{fig:AES_n=2_Useries}
\end{figure}

Let us start the discussion of the results with the simple case of the completely filled band: $n=2$. 
Creating two holes with opposite spins constitutes a two-body problem and a closed analytical expression for the excitation spectrum is readily derived. \cite{Cin77,Saw77,Nol90}
Alternatively, one may apply the Chebyshev expansion method using exact states for rather large systems, since the Hilbert-space dimension is merely $L^2$. 
Some numerical results for $L=1000$ and periodic boundary conditions are shown in Fig.\ \ref{fig:AES_n=2_Useries}.

The two-hole spectra for different $U$ can be understood as follows: 
If $U=0$, one can straightforwardly evaluate Eq.\ (\ref{eq:A2hole}) by Fourier transformation to reciprocal space. 
This yields
\be
A^{(0)}_{\rm 2-hole}\lr{\omega} 
= 
\int dx \,
\rho_0(x)
\rho_0 (\omega-x)
\: , 
\ee
where $\rho_0(x) = L^{-1} \sum_{k} \delta(x-\varepsilon(k))$ and $\varepsilon(k) = - 2 t \cos(k)$, i.e., the two-hole spectrum is given by the self-convolution of the non-interacting density of states. \cite{L53}
The self-convolution of the one-dimensional density of states $\rho_{0}(x)$ is just given by the two-dimensional one with bandwidth $2W=8t$ (see black line).
The two conduction-band holes move independently of each other through the lattice. 

Upon increasing $U$, a sharp satellite emerges at the lower edge of this ``band-like part'' of width $2W$ and moves to higher binding energies.\cite{Cin77,Saw77}
For $U \gg t$ it is approximately separated by $U$ from the barycenter of the band-like part located at $\omega=-4t$. 
This corresponds to the formation of a stable two-hole bound state which we will call ``doublon'' in the rest of the paper.
Since the one-hole spectrum has a finite support, a decay of the doublon into independent holes is prohibited by energy conservation for large $U$. 
One dimension is in fact special in that there is a bound two-hole state outside the continuum for any $U>0$, so that momentum conservation must be considered as well to understand its stability. \cite{HDD06}
This is very similar to the Bose case discussed in Ref.\ \onlinecite{WTL+06}.

\begin{figure}[b]
\includegraphics[width=0.75\columnwidth]{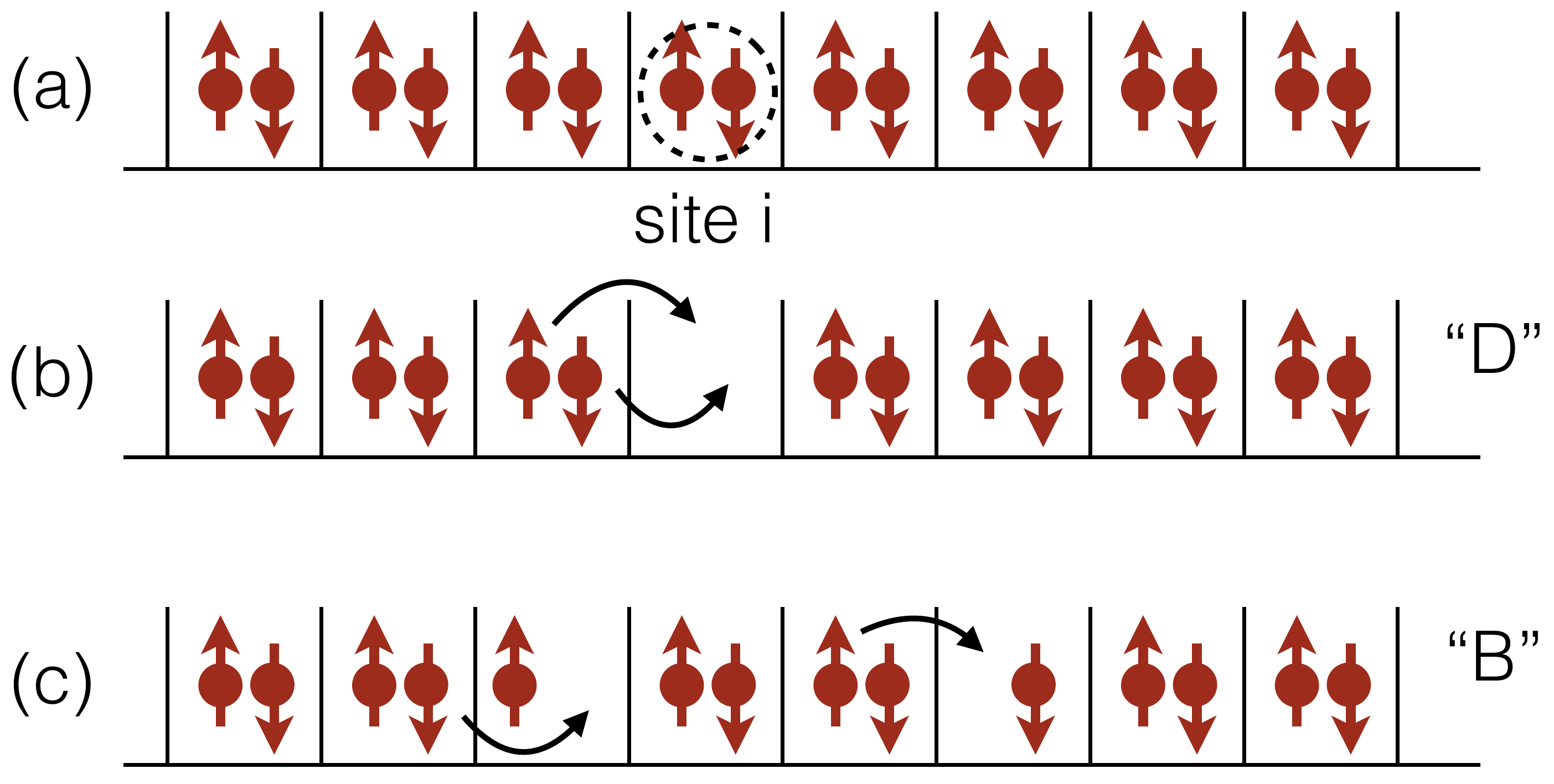}
\caption{
(a) Ground state $\ket{0,N}$ for $N=2L$ electrons (filling $n=2$).
The two-hole spectroscopy removes two electrons at site $i$ (dashed line).
(b) Configuration of an excited state $\ket{n,N-2}$ contributing to the doublon satellite (``D''). 
Arrows indicate doublon propagation. 
(c) Configuration contributing to the band-like part (``B'').
Arrows indicate independent motion of the two final-state holes.
}
\label{fig:db}
\end{figure}

The finite width of the satellite reflects the fact that the doublon is itinerant. 
For strong $U$ the propagation of a single doublon is perturbatively described by a second-order hopping process and is captured by the effective Hamiltonian (see also Eq.\ (\ref{eq:Heff}) below)
\begin{equation}
  H_{\rm eff} 
  = 
  \frac{2t^{2}}{U} \sum_{\langle ij \rangle} \lr{d_{i}^{\dagger} d_{j} + \mbox{H.c.}} + \lr{\frac{4t^{2}}{U}+U} \sum_i n_i^d 
  \: ,
\end{equation}
where $d^\dagger_i = c^{\dagger}_{i\uparrow}c^{\dagger}_{i\downarrow}$ is a bosonic doublon creator and $n_i^d = d^\dagger_i d_i$. 
Hence, in the strong-coupling limit the satellite has the shape of the non-interacting density of states with a rescaled bandwidth of $8t^{2}/U$. 
With increasing $U$, the doublon satellite gains more and more spectral weight at the expense of the band-like part and eventually results in a $\delta$-peak for $U=\infty$, corresponding to a completely immobile doublon.

In the strong-$U$ limit, the physics can be visualized to some extent by typical electronic configurations contributing to the different spectral features.
Fig.\ \ref{fig:db} gives a sketch of the different configurations contributing to the doublon satellite ``D'' and to the band-like part ``B''.

\subsection{Partially filled band: Overview}
\label{sec:part}

\begin{figure*}
\includegraphics[width=1.5\columnwidth]{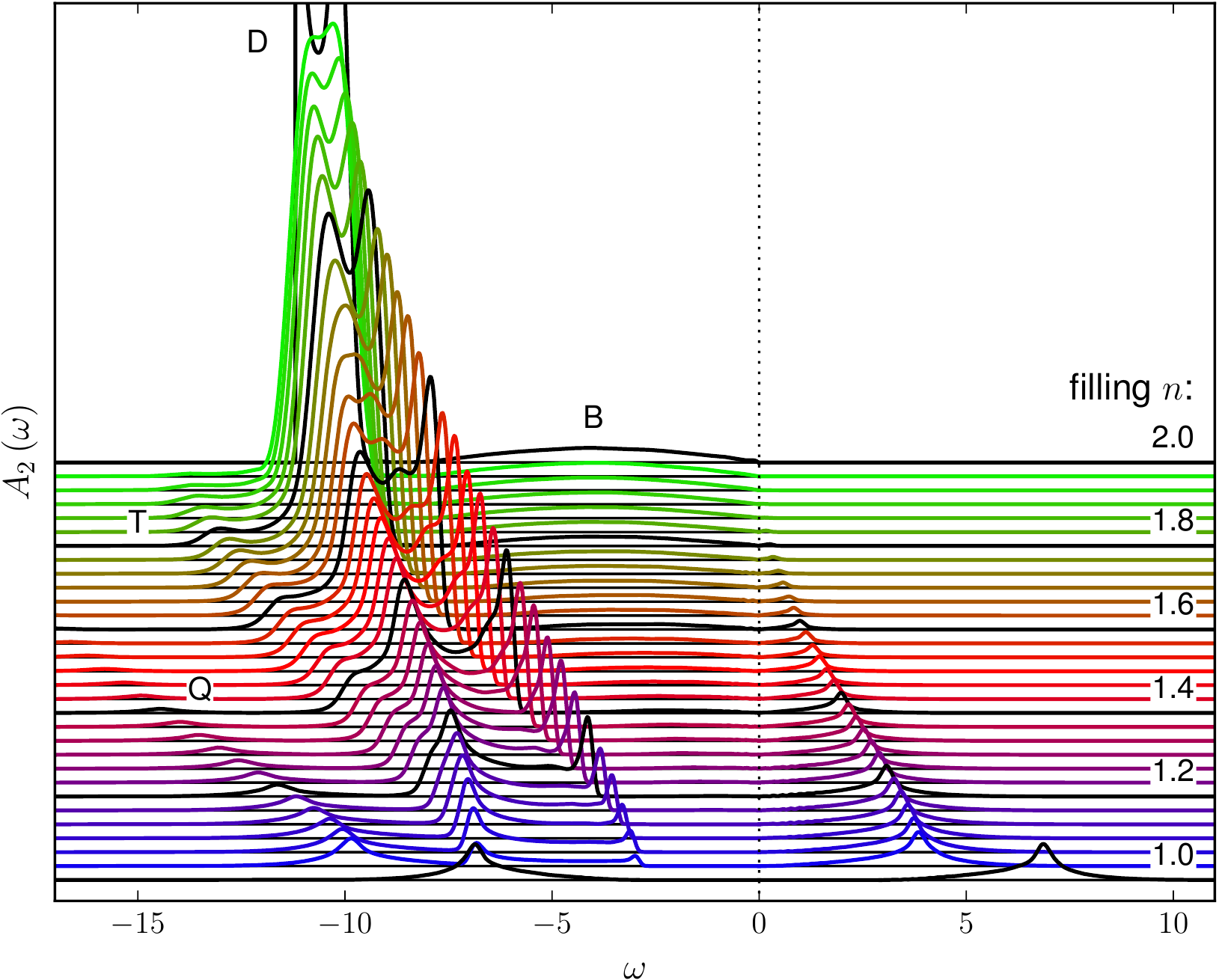}
\caption{
Two-hole excitation spectrum ($\omega<0$) and two-particle excitation spectrum ($\omega>0$) of the one-dimensional  Hubbard model at $U=6t$ for fillings $n$ between half-filling ($n=1$) and the limit of the completely filled band ($n=2$).
The filling values indicated on the right are drawn in black.
Results as obtained by DMRG for a system with $L=60$ lattice sites and open boundary conditions at an energy resolution of $\delta E = 0.2t$, corresponding to up to $M=1149$ Chebyshev moments.
The main spectral structures are labeled by ``D'' (doublon), ``B'' (band-like part), ``T'' (triplon) and ``Q'' (quadruplon) (see text).
}
\label{fig:nseries}
\end{figure*}

We now turn to the case of a conduction band with a finite hole density: $n<2$.
This changes the situation completely as the two additional holes placed with opposite spins at site $i$ interact with a thermodynamically relevant fraction of holes in the incompletely filled band. 
From the technical point of view, the computation of the spectrum transmutes into a severe many-body problem: 
A comprehensive and reliable study, however, is possible when tolerating a non-zero low-energy cutoff in the form of a finite spectral resolution $\delta E$. 
Using the Chebyshev technique combined with matrix-product states, we have been able to study the excitation spectra in the entire filling range $1 \le n \le 2$ for the Hubbard model on lattices as large as $L=60$ at a constant energy resolution $\delta E \approx 0.2t$. 
With $\delta E \lesssim 0.2t$ one would start to resolve finite-size artifacts.
Calculations have been performed for systems with open boundary conditions and choosing $i$ as a central site in the chain [see Eqs.\ (\ref{eq:A2hole}) and (\ref{eq:A2particle})].

Figure \ref{fig:nseries} provides an overview and shows the filling dependence of the spectrum at $U=6t$, chosen such that for $n=2$, the doublon satellite is clearly separated from the band-like part. 
Several distinct effects arise at finite hole density, which we discuss now in turn:

First of all, the overall intensity of the two-hole spectrum in the range $\omega < 0$ is seen to decrease with decreasing filling. 
This is a consequence of the sum rule for the double occupancy, 
\be
  \avg{n_{i\uparrow}n_{i\downarrow}} = \int_{-\infty}^{0} d\omega~ A_{\rm 2-hole}\lr{\omega} \: , 
\label{eq:sumrule}
\ee
which is straightforwardly derived from Eq.\ (\ref{eq:A2hole}), and of the fact that $\avg{n_{i\uparrow}n_{i\downarrow}}$ is decreasing with $n$. 

The doublon satellite (``D'') is still present with strong weight at $U=6t$ but its support broadens significantly with decreasing $n$. 
The satellite can be traced down to fillings close to half-filling. 
Right at $n=1$ it vanishes completely. 
An extended discussion is given in Sec.\ \ref{sec:doublon}.

A continuation of the ``band-like part'' (``B'') of the two-hole spectrum is also clearly visible, at least down to $n\approx 1.5$. 
When the filling approaches half-filling, however, the weight of the band-like part becomes very small, and the structure is hardly visible at low energies, say $-2 \lesssim \omega <0$, for fillings $1 < n < 1.2$. 
Note, however, that there is finite spectral weight for $\omega \to 0$ at any filling $n>1$;
only at half-filling, $n=1$, is the spectrum gapped.

As soon as $n<2$, the spectrum exhibits an additional structure, marked as ``T'' in Fig.\ \ref{fig:nseries}.
For fillings $n \to 2$ it shows up as a tiny peak at high binding energies. 
It gains some spectral weight upon decreasing $n$ and moves towards $\omega=0$, until it completely merges with the doublon satellite. 
At the same time, another peak can be seen emerging within the spectral range of the doublon satellite (seen best for $n=1.6$ at $\omega \approx -8.7t$). 
It moves towards $\omega=0$ at the same rate and also belongs to ``T''. 
The structure ``T'' will be interpreted as a ``triplon'' in Sec.\ \ref{sec:triplon}.

The physics close to and at half-filling is complicated by the appearance of yet another structure ``Q'' which emerges for $n \lesssim 1.5$ at the highest binding energies and gains more and more weight as $n$ approaches half-filling. 
This feature will be addressed in Sec.\ \ref{sec:quad} and interpreted as a ``quadruplon''.

At half-filling the system is a Mott insulator. 
The gap in the one-particle spectral function is accessible to the Bethe-ansatz method \cite{LW68} and can be computed from the exact expression \cite{Ovc69}
\begin{equation}
  \Delta = \frac{16t^2}{U} \int_1^\infty dx \: \frac{\sqrt{x^2-1}}{\sinh(2\pi x ~ t / U)} \: .
\label{eq:gap}
\end{equation}
At $U=6t$ it amounts to $\Delta = \mu_{+} - \mu_{-} \approx 2.89t$. 
As Fig.\ \ref{fig:nseries} demonstrates, the two-particle spectral function is gapped as well, but by twice the one-particle gap $2 \Delta \approx 5.86t$ (the numerical value is obtained from the ground-state energies, and the slight discrepancy compared to twice of the Bethe-ansatz value is due to the finite system size).

Note that the chemical potential jumps from $\mu = \mu_{-}$ for $n \nearrow1$ to $\mu = \mu_{+}$ for $n \searrow 1$ and is not uniquely defined at half-filling. \cite{mu}
The two-hole and the two-particle spectrum for $n=1$ in Fig.\ \ref{fig:nseries} have been calculated for the choice $\mu = (\mu_{+} + \mu_{-})/2 = U/2 = 3t$, which implies a symmetric spectrum $A_{2}(-\omega) = A_{2}(\omega)$.
If $\mu = \mu_{+}$ had been chosen, the $n=1$ spectrum would have evolved continuously from the spectrum for $n>1$. 
Hence, the single broad peak in the two-hole spectrum at $n=1$, which takes the whole spectral weight, smoothly connects to the quadruplon peak ``Q'' at higher fillings.

For fillings below half-filling, $n<1$, the two-hole spectrum is related to the two-particle spectrum for $n>1$ via the relation Eq.\ (\ref{eq:phsym}).
As is seen for $\omega>0$ in Fig.\ \ref{fig:nseries}, it consists of a single peak, smoothly connected to ``Q'' which just continues to lose spectral weight until it vanishes for $n\to 0$.
The vanishing of the weight is related to sum rule for the two-particle spectrum
\be
  \avg{(1-n_{i\uparrow}) (1-n_{i\downarrow})} = \int_{0}^{\infty} d\omega~ A_{\rm 2-particle} \lr{\omega} \: , 
\label{eq:sumrulep}
\ee
which is straightforwardly derived from Eq.\ (\ref{eq:A2particle}), and to the fact that the average number of empty sites goes to zero for $n \to 2$, i.e., the phase space for creating two additional {\em particles} at a site $i$ vanishes.

Let us also mention that the {\em difference} between the total weights of the two-hole and the two-particle spectrum is given by the filling: 
\be
  \int_{-\infty}^{0} d\omega~ A_{\rm 2-hole} \lr{\omega} 
  -
  \int_{0}^{\infty} d\omega~ A_{\rm 2-particle} \lr{\omega} 
  = n-1 \: .
\ee
This is a direct consequence of Eqs.\ (\ref{eq:sumrule}) and (\ref{eq:sumrulep}).

\subsection{The doublon}
\label{sec:doublon}

For fillings close to the simple $n=2$ limit, the physical interpretation of the doublon peak ``D'' as a repulsively bound pair of holes is still self-evident. 
With decreasing $n$, and for fillings close to half-filling in particular, the validity of this picture must be questioned seriously.

First of all, as soon as $n<2$, the doublon starts to interact with a continuum of electron-hole excitations, such that the corresponding peak should actually be seen as a resonance with a finite lifetime rather than an exact eigenstate of the Hamiltonian (for a given wave vector). 
This will be discussed in detail in Sec.\ \ref{sec:res:kdependence}.

In fact, the doublon peak broadens with decreasing $n$. 
Its support widens from a value slightly larger than $\Delta \omega = 8 t^{2} / U = 4t/3$, which is the perturbative result for $U\to \infty$ at $n=2$, to $\Delta \omega \gtrsim 4t$ for fillings close to, but above half-filling.

\begin{figure}[b]
\includegraphics[width=0.75\columnwidth]{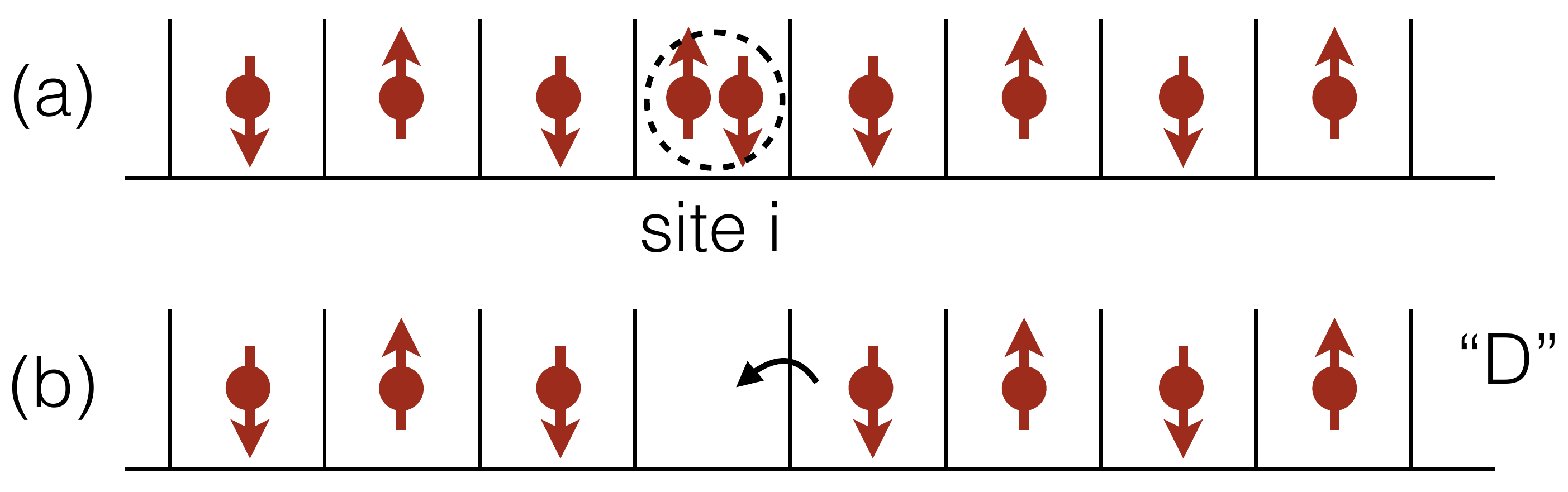}
\caption{
(a) Configuration with $N=L+1$ electrons contributing to the ground state $\ket{0,N}$. 
Two electrons at site $i$ are removed (dashed line).
(b) Configuration contributing to an excited state $\ket{n,N-2}$ related to the doublon satellite (``D''). 
The arrow indicates doublon propagation. 
}
\label{fig:db2}
\end{figure}

To understand the main reason for this broadening, consider the spectrum for a system with $N=L+1$ particles, i.e., for a filling slightly above half-filling $n=1$. 
Fig.\ \ref{fig:db2} provides a sketch. 
The support of the doublon peak is given by the spectral range of the final states. 
As indicated in the sketch, the dominating final-state configurations are characterized by a hole moving through a lattice of half-filled sites. 
This is a standard many-body problem, intensively discussed, e.g., in the context of the Mott-Hubbard metal-insulator transition:\cite{Geb97}
Double occupancies are effectively suppressed at strong $U$, well-formed local magnetic moments emerge and develop strong antiferromagnetic correlations due to the Anderson super-exchange mechanism.
While the effective coupling constant $J=4t^{2} / U$ of the latter is derived in second-order-in-$t$ perturbation theory and is small, the motion of the hole takes places on the large energy scale $t$.
Neglecting $J$ altogether, already provides the right width of about $4t$ for the doublon peak, consistent with the corresponding spectrum displayed in Fig.\ \ref{fig:nseries}. 

Let us mention that the arguments can be formalized by considering the standard mapping of the low-energy sector of the Hubbard model for strong $U$ onto the $t$-$J$ model. \cite{CSO78} Close to half-filling it reads:
\begin{equation}
  H_{t-J} 
  = 
  - t \sum_{\langle ij \rangle, \sigma} \lr{ \widetilde{c}^{\dagger}_{i\sigma} \widetilde{c}_{j\sigma} + \mbox{H.c.}}
  + J \sum_{\langle ij \rangle} \ff S_{i} \ff S_{j}
  \: , 
\label{eq:tj}
\end{equation}
where
$\ff S_{i} = \frac12 \sum_{\sigma\sigma'} c_{i\sigma}^{\dagger} \ff \sigma_{\sigma\sigma'} c_{i\sigma'}$ is the spin at site $i$, $\ff \sigma$ the vector of Pauli matrices
and where
$\widetilde{c}_{i\sigma}$ is the annihilator in the subspace with no double occupancies. 
Neglecting $J$ yields the $t$ model which exactly maps \cite{Lac85} onto the $J_{\rm KL}=\infty$ Kondo lattice. 
For a single hole in the $t$ model or, equivalently, a single electron in the Kondo lattice, the ground state is a ferromagnet as is known from Nagaoka's work \cite{Nag66} or, equivalently, can be proven \cite{TSU97b} for the Kondo lattice. 
The width $4t$ is obvious then. 
For the present case with finite $U$, a somewhat larger width is expected. 

Another helpful observation is that the final states of the two-hole spectroscopy (see Fig.\ \ref{fig:db}b) are the same as the final states of the one-hole spectroscopy, i.e., photoemission, but from an exactly half-filled conduction band with $N=L$ electrons. 
The (raw) photoemission spectrum is given by the one-hole spectral function
\begin{eqnarray}
A_{{\rm 1-hole},\sigma}\lr{\omega} 
&=& 
\sum_n \norm{\matrixel{n,N-1}{c_{i\sigma}}{0,N}} 
\nonumber \\
&\times&
\delta\lr{\omega + \mu - (E_0^{\lr{N}}-E_n^{\lr{N-1}})}
\: . 
\label{eq:A1hole}
\end{eqnarray}
This shows that the support of the doublon peak must be the same as the support of the photoemission spectrum. 
The latter has indeed a width of $\Delta \omega \gtrsim 4t$ at $U=6t$, as can be read off from the DMRG results shown in Fig.\ \ref{fig:pes} of Appendix \ref{sec:pes}.
As the matrix elements for the two-hole spectroscopy are different, however, the shape of the doublon satellite is different and, at least on this level, it is difficult to make contact with the well-known spinon and holon excitations that govern the one-hole spectrum. 

We conclude that the concept of a doublon as a repulsively bound pair of holes breaks down for fillings close to half-filling. 
The main difference is that the doublon moves in a background of {\em singly} occupied sites and explores antiferromagnetic correlations. 
This strongly affects its mobility. 
For $n \to 1$ the doublon propagates on the energy scale of about $4t$, opposed to the energy scale $8t^{2}/U$ set by  double-hopping processes in the high-density limit. 
The strong decrease of the weight of the doublon peak with decreasing $n$ is related to the decreasing density of doubly occupied sites in the respective initial ground state.
The shift of the position of the doublon satellite is closely related to the filling dependence of the band-like part.

\subsection{The band-like part}
\label{sec:band}

The band-like part ``B'' of the two-hole spectrum results from the continuum of unbound two-hole final states. 
For $n=2$ and $U=0$, this is given by the self-convolution of the density of states $\rho_{0}(x)$. 
With increasing $U$ it quickly loses weight in favor of the doublon satellite (see Fig.\ \ref{fig:db}) and becomes an almost featureless structure for strong $U$ whose shape still loosely resembles the self-convolution of the density of states, but without the strong van Hove singularities.
Its weight is already low for $U = 6t > W$, see Fig.\ \ref{fig:AES_n=2_Useries}. 

For fillings close to half-filling, this scenario still holds to a good approximation, but with the density of states replaced by its occupied part only (for a given filling).
Since the occupied fraction shrinks with decreasing $n$, the weight of the band-like part must diminish and its spectral support narrow down, with the barycenter shifting upwards.
This effect is clearly visible in Fig.\ \ref{fig:nseries} for high fillings.

The barycenter of the band-like part also fixes the position of the doublon satellite which is found at an energy of about $U$ below.
Assuming a rectangular density of states with bandwidth $W=4t$, we have $\omega_{0}=-(n/2) W$ for the barycenter and $\Delta \omega =n W$ for the width of the band-like part. 
This simple model roughly explains the trend seen for fillings $n \gtrsim 1.6$.
For example, the model predicts $\omega_{0} = -3.2t$ for $n=1.6$ resulting in $\omega = -(n/2) W - U = - 9.2$ for the satellite position. 
This should be compared with the barycenter of the satellite at $\omega \approx -8.75t$ in Fig.\ \ref{fig:nseries} (obtained by integrating the data).

For fillings $n \lesssim 1.5$, however, $\omega_{0}$ shifts much faster with decreasing $n$ and the model breaks down (note that there is a finite band-like part for any $n>1$).
In this filling range, one may consider the two additional holes in the final state as moving independently of each other, but strongly interacting with the high hole density in the ground state.

This can be very roughly described in the following way: 
We replace the non-interaction density of states $\rho_{0}$ by an interacting density of states $\rho$ which is assumed to consist of two Hubbard peaks, separated in energy by $U$. 
For $n \ge 1$ and following the standard theory \cite{MES93} of filling-dependent spectral-weight transfer in the strong-coupling limit, the lower Hubbard band has the weight $2-n$ and is fully occupied, while the upper Hubbard band is partially occupied with the occupied fraction having the weight $2n-2$ (and the unoccupied fraction having weight $2-n$).
In the simple model the band-like part is then proportional to the self-convolution of the low-energy part of the occupied density of states $\rho$, i.e., of the occupied part of the upper Hubbard band.
Assuming that the shape of the latter is $n$-independent, we find $\omega_{0} = - (n-1) W$.

This roughly explains the filling dependence of the doublon satellite for $n \lesssim 1.6t$.
For example, we have $\omega_{0} = -0.8t$ for $n=1.2$, resulting in $\omega = - (n-1) W - U = - 6.8t$ for the satellite position. 
This compares well with the DMRG results in Fig.\ \ref{fig:nseries} where we find $\omega \approx -6.3t$.
For $n \to 1$, in particular, the model predicts a vanishing band-like part with $\omega_0 \to 0$, and the resulting position of the doublon satellite at $\omega \approx - U$ agrees well with the data.

Although the model is surely oversimplified, it illustrates that the filling-dependent spectral-weight transfer is important for the location of the continuum of unbound two-hole final states, and thus for the doublon satellite as well. 

\subsection{The triplon}
\label{sec:triplon}

\begin{figure}[b]
\includegraphics[width=0.75\columnwidth]{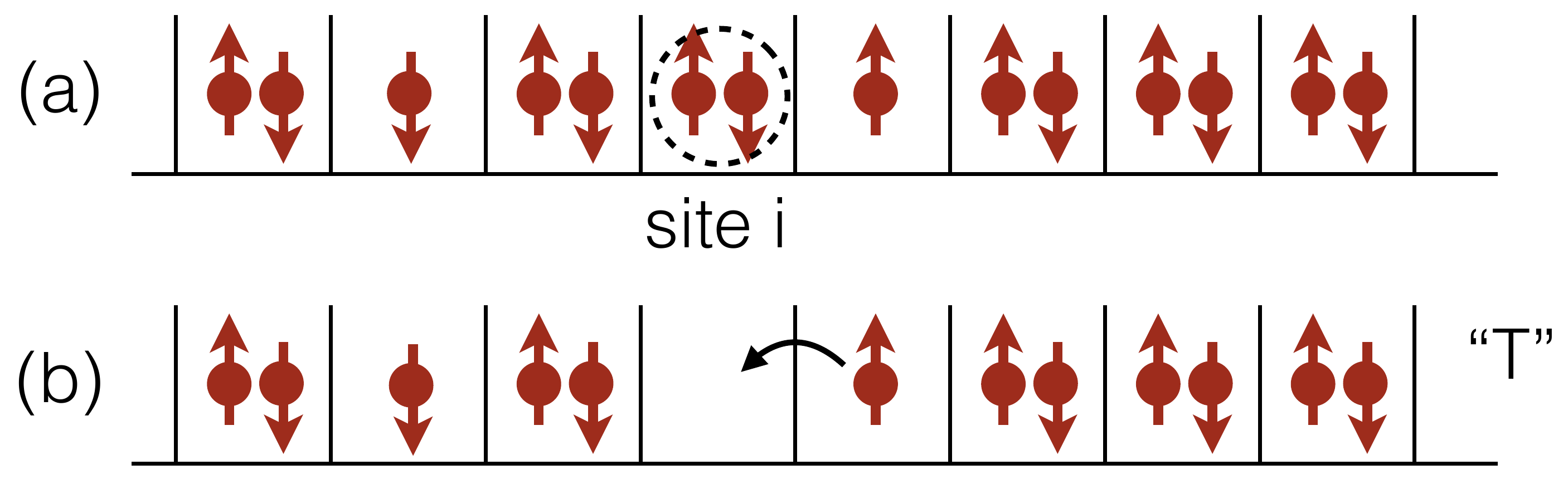}
\caption{
(a) Ground state $\ket{0,N}$ for $N=2L-2$ electrons ($n<2$, high filling).
Two electrons at site $i$ are removed (dashed line).
(b) Configuration of an excited state $\ket{n,N-2}$ contributing to the triplon satellite (``T''). 
The arrow indicates the internal degree of freedom of the triplon.
}
\label{fig:triplon}
\end{figure}

\begin{figure}[t]
\includegraphics[width=0.95\columnwidth]{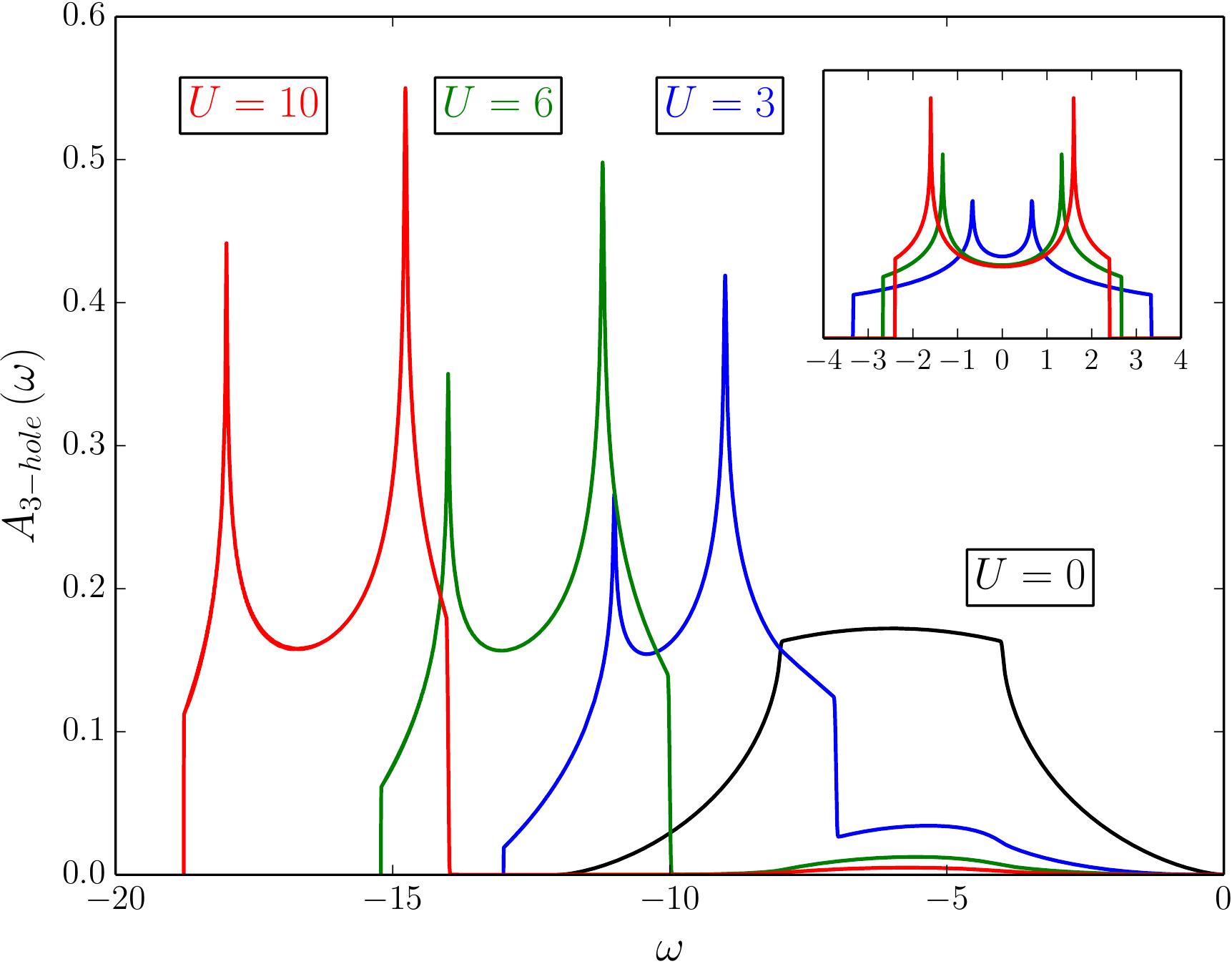}
\caption{
Three-hole excitation spectrum [see Eq.\ (\ref{eq:threehole})] for $L=500$ sites, different $U$ and $n=2$. 
Calculations are performed at $\delta E = 0.01t$ (up to $M=1876$ Chebyshev moments), using exact states and periodic boundary conditions. 
The inset shows the convolutions $\int dx~ \rho_0\lr{\omega-x,4t} \rho_0 \lr{x,8t^2/U}$ for the same values of $U$, where $\rho_0 \lr{x,W} \equiv 1 / \pi \sqrt{W^2/4-x^2}$.
}
\label{fig:tri2}
\end{figure}

The most apparent deviations from the standard Cini-Sawatzky theory consist in the appearance of the new structures ``T'' and ``Q''.
Let us first concentrate on ``T'' which already shows up for low hole-densities $n\lesssim2$, a limit which has been studied to some extent using the ladder approximation in the hole-hole channel. \cite{Cin79,DK84}

When acting with $c_{i\uparrow}c_{i\downarrow}$ on the ground state, one may distinguish between contributions to the spectrum resulting from two different types of ground-state configurations:
In the case of an excitation that takes place in a hole-free region, one expects essentially the same shape of the spectrum as in the $n=2$ case.
For $n \to 2$ this is the dominant contribution. 
If the excitation takes place at a site $i$ next to an already existing hole, however, a final state with three holes on nearest-neighboring sites is created.
This case is schematically illustrated with Fig.\ \ref{fig:triplon}.

Here, the question arises whether the three holes can form a new bound object, a ``triplon''. 
Since this question addresses the excited states one projects onto in Eq.\ (\ref{eq:A2hole}), rather than the ground state, we first study the three-hole excitation spectrum
\begin{eqnarray}
A_{\rm 3-hole,\sigma} \lr{\omega} 
&=& 
\sum_n \norm{
\matrixel{n, N-3} {c_{i\uparrow}c_{i\downarrow}c_{i+1\sigma}} {0,N}
} 
\nonumber \\
&\times& 
\delta \lr{
\omega + 3 \mu -\lr{E_0^{\lr{N}}-E_n^{\lr{N-3}}}
}
\label{eq:threehole}
\end{eqnarray}
for the completely filled band ($n=2$). 
This involves the same type of final states while the ground state is trivial, and we are left with a much simpler three-body problem only.
Since the Hilbert-space dimension is approximately $L^3/2$, we can use the Chebyshev expansion method with exact states.

The result for $L=500$ sites is shown in Fig.\ \ref{fig:tri2}. 
One observes a similar general pattern as in the two-hole case: 
For $U=0$ the spectrum consists of a three-fold self-convolution of $\rho_0(\omega)$ with a spectral width given by $3W=12t$, while for finite $U>0$, there is a band-like part and a split-off correlation satellite.

For finite but strong $U$, the satellite in $A_{\rm 3-hole,\sigma}\lr{\omega}$ has a slightly asymmetric weight distribution due to the presence of the band-like part. 
Apart from that, its shape is given by a convolution of $\rho_0(\omega)$ (the single hole, width $W$) with the same density of states, but rescaled (the doublon, width $\Delta \omega = 8 t^2/U$), see the inset in Fig.\ \ref{fig:tri2}. 
This clearly indicates that the doublon and the hole propagate independently and do not form a bound state.

This should also hold for the two-hole spectrum (Fig.\ \ref{fig:nseries}): 
The structure ``T'' does not represent a stable triplon, but is rather a doublon-hole continuum superimposed on the doublon satellite. 
The filter-operator analysis in Sec.\ \ref{sec:filter} and further arguments given below indeed support this interpretation.

Compared to the three-hole spectrum, ``T'' in Fig.\ \ref{fig:nseries} has smaller spectral weight, but the same shape. 
We would like to stress that the two peaks belonging to ``T'' are simply a consequence of the van Hove singularities in one dimension, separated by about $16t^{2}/ U \approx 2.67 t$ for $U=6t$, and should not be mistaken for separate resonances.
Hence, they will not be observed for higher-dimensional lattices.

\begin{figure}[t]
\includegraphics[width=0.95\columnwidth]{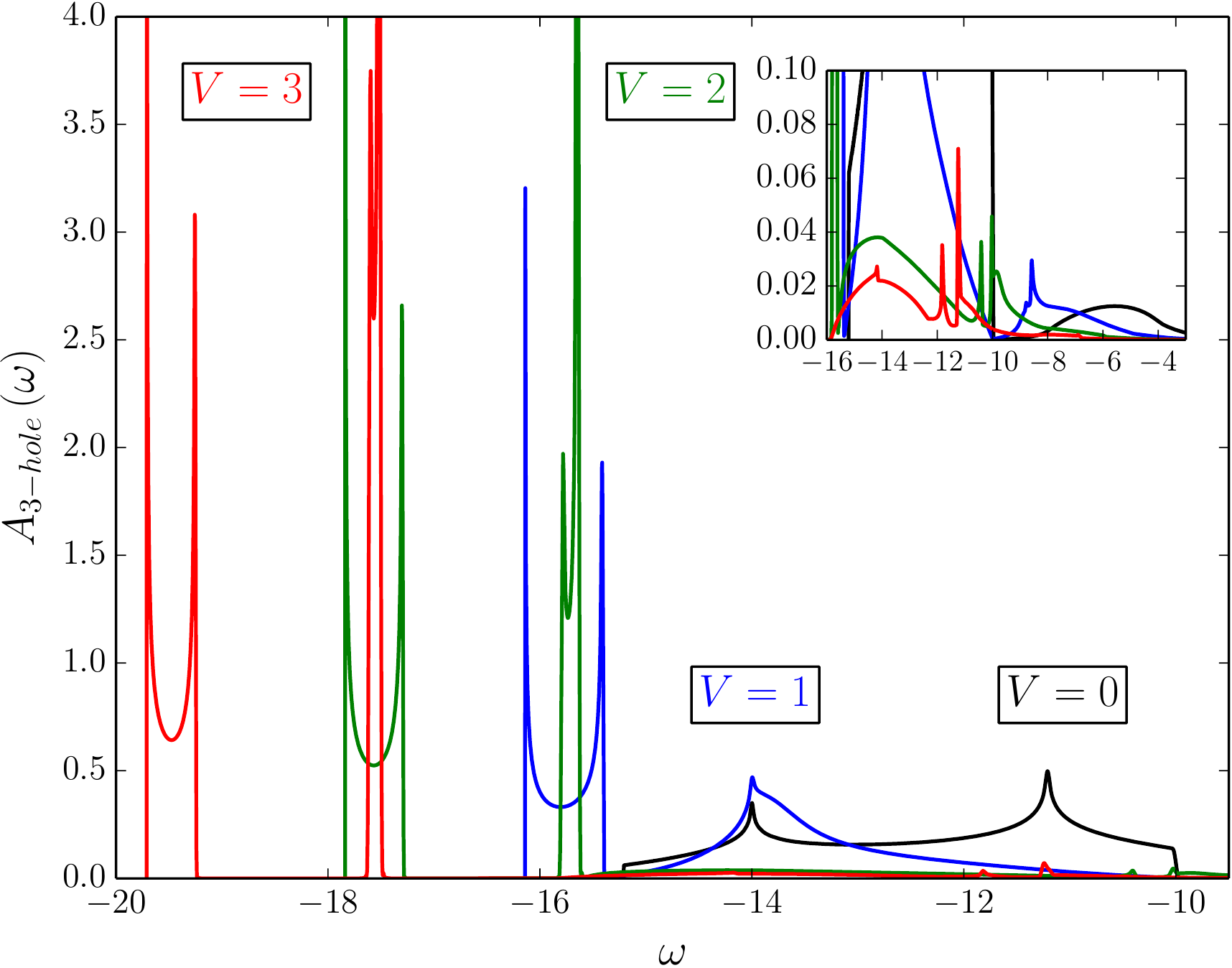}
\caption{
Three-hole excitation spectrum as in Fig.\ \ref{fig:tri2}, but for different values of the nearest-neighbor interaction $V$ and fixed $U=6t$ (up to $M=1971$ Chebyshev moments). The inset shows a close-up view of the continua.
}
\label{fig:tri2v}
\end{figure}

The interpretation of ``T'' as a doublon-hole continuum also explains that it is located at approximately $U$ below the barycenter of the three-hole continuum. 
The latter is given by $\omega_{0}=-6t$ for $n=2$. 
With decreasing filling the support of the three-hole continuum shrinks, and so does its barycenter, but at a faster rate than the two-hole continuum, so that the doublon-hole continuum ``T'' eventually merges with the doublon satellite for fillings approaching half-filling.

A stable, repulsively bound triplon forms in the presence of an additional repulsive nearest-neighbor Coulomb interaction $V$. 
This situation is shown in Fig.\ \ref{fig:tri2v}. 
Starting from the doublon-hole continuum at $U=6t$ and $V=0$, and switching on $V$ leads to the emergence of two satellites splitting off for sufficiently strong $V$.
Both have the same shape as the non-interacting density of states $\rho_{0}(x)$ and are separated by about $2V$ from the continuum. 
The presence of two satellites rather than one is due to the fact that a triplon has an internal degree of freedom: 
On two sites, the three-hole states 
$(1,0)^{T} \equiv c^{\dagger}_{1\uparrow} | {\rm vac.} \rangle$ and
$(0,1)^{T} \equiv c^{\dagger}_{2\uparrow} | {\rm vac.} \rangle$ 
form two configurations. 
When coupled by the Hamiltonian
\be
H_{2} 
= 
\left(
\begin{array}{cc}
U+2V & -t \\
-t & U+2V
\end{array}
\right) \; ,
\label{eq:triinternal}
\ee
they result in a pair of bonding/antibonding eigenstates at energies $E_{\pm} = U + 2V \pm t$.
Note that with $H_{2}$, we count the interaction between holes rather than electrons. 
The energy splitting of $2t$ very well explains the energy difference between the positions of the two satellites in the spectra in Fig.\ \ref{fig:tri2v}.
Their widths are different, but become equal in the strong-coupling limit $U,V \gg t$ and are then given by $8 t^{2} / (U+2V)$.

\begin{figure}[t]
\includegraphics[width=0.95\columnwidth]{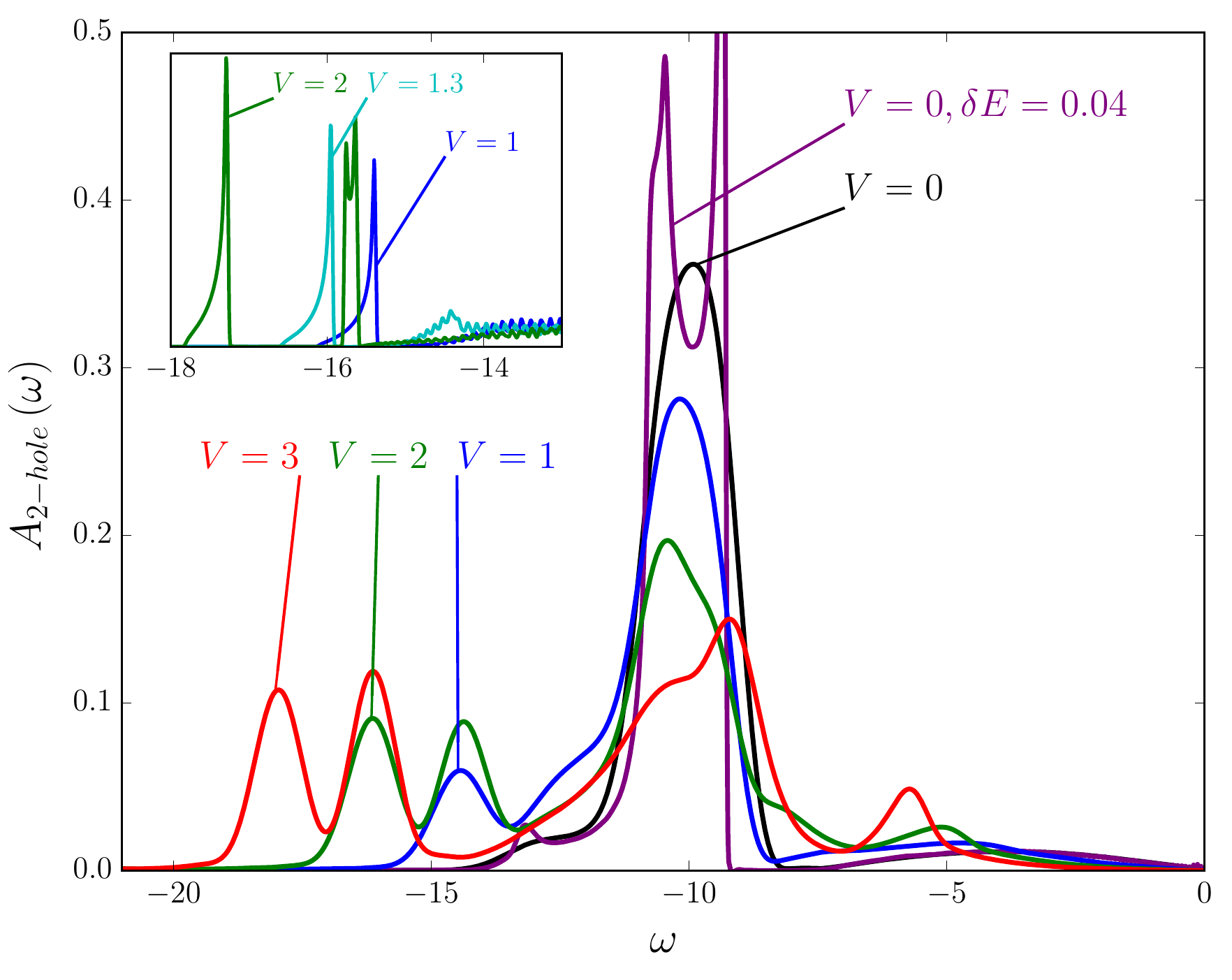}
\caption{Two-hole excitation spectrum at filling $n=1.8$ for $U=6t$ and different $V$ as indicated.
DMRG calculations for $L=60$ lattice sites and
energy resolution $\delta E = 0.4t$ (up to $M=376$ Chebyshev moments), except for the purple line at $V=0$ ($L=120$, $\delta E=0.04t$, $M=3553$).
Inset: spectrum at high binding energies for $L=100$ and $n=1.98$. Calculations for $\delta E=0.01t$ (up to $M=2852$) using exact states and periodic boundary conditions.
}
\label{fig:aesv}
\end{figure}

Fig.\ \ref{fig:aesv} shows the emergence of the two triplon satellites in the two-hole spectrum at $n=1.8$ as obtained by DMRG.
With $V > 0$, we find a much stronger growth of the entanglement entropy during the Chebyshev iteration.
This is to be expected, as the model becomes non-integrable.\cite{MURF11}
These spectra have therefore been calculated at a moderate energy resolution $\delta E = 0.4t$. 
For $V=3$, for example, the MPS bond dimension at this point already exceeds $6200$.

For $V=0$, we find the dominating doublon satellite around $\omega \approx -10t$ and the two peaks of the doublon-hole continuum at higher binding energies, separated by about $16t^{2} / U \approx 2.67t$. 
One is located at $\omega \approx - 13.1 t$, while to other one is next to the intense van Hove singularity of the doublon structure at $\omega \approx - 10.4 t$, and barely distinguishable from it. 
It requires a high-resolution calculation to uncover this feature ($\delta E = 0.04t$, the purple curve in Fig.\ \ref{fig:aesv}).
For $V>0$, the double-peak structure of the triplon appears, with a separation of $2t$ and shifting by about $2V$ with increasing $V$, as in the case of the three-hole spectrum (Fig.\ \ref{fig:tri2v}).
The doublon satellite, on the other hand, shows a considerable loss of spectral weight, but has a basically unchanged position when increasing $V$.
In addition, we observe another resonance emerging out of the band-like part at $\omega\approx -4t-V$, which we can identify as two neighboring holes repulsively bound by $V$ as a dimer.

The available resolution does not allow us to determine the fine structure of the triplon peaks. 
The linear-prediction technique, \cite{Sch11,GTVHE14,WCPS14} which is commonly used to enhance the resolution, has not provided us with additional spectral information in this case. 
Instead, we go to an even higher filling, $n=1.98$, where an exact calculation is still feasible for a fairly large system. 
The result for $L=100$ and $\delta E=0.01$ is shown in the inset of Fig.\ \ref{fig:aesv}. 
We observe that while the first of the triplon peaks appears right away, the second one only rises above the continuum when $V$ exceeds a critical value of $V_c \approx 1.3$. 
Note that the shape of the former markedly differs from the free density of states, indicating that in contrast to the three-hole spectral function (Fig.\ \ref{fig:tri2v}), the triplon in $A_{\rm 2-hole,\sigma}\lr{\omega}$ does not become stable in the whole Brillouin zone.

\subsection{The quadruplon}
\label{sec:quad}

\begin{figure}[b]
\includegraphics[width=0.75\columnwidth]{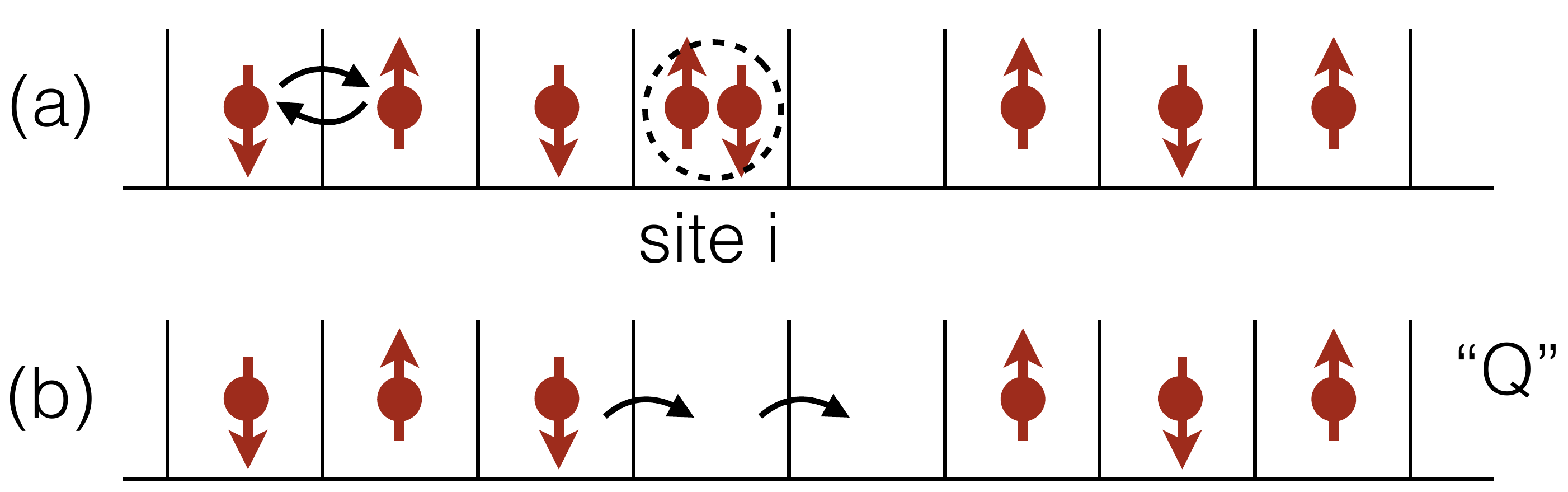}
\caption{
(a) Configuration with $N=L$ electrons (half-filling $n=1$) contributing to the ground state $\ket{0,N}$. 
Two electrons at site $i$ are removed (dashed line).
Arrows: virtual hopping processes in second-order-in-$t$ perturbation theory. 
(b) Configuration of an excited state $\ket{n,N-2}$ contributing to the quadruplon satellite (``Q''). 
Arrows indicate quadruplon propagation. 
}
\label{fig:quad1}
\end{figure}

The two-hole spectrum of the half-filled Hubbard model (see Fig.\ \ref{fig:nseries}, lowest panel) displays a broad peak ``Q'' centered around $\omega \approx -7.4t$. 

A two-hole excitation requires a double occupancy at a site $i$ in the ground state $\ket{0,N}$.
For strong $U$ and $N=L$, a corresponding configuration is generated by a virtual second-order hopping process, which at the same time favors antiferromagnetic spin correlations. 
This is the famous Anderson super-exchange. \cite{And59}
Starting from this initial-state configuration, the resulting configuration contributing to the $N-2$-electron final state $\ket{n,N-2}$ must therefore contain two nearest-neighboring empty sites, i.e., two neighboring doublons (see the sketch given in Fig.\ \ref{fig:quad1}).
One may ask whether these bind to form a heavier ``quadruplon''.

\begin{figure}[t]
\includegraphics[width=\columnwidth]{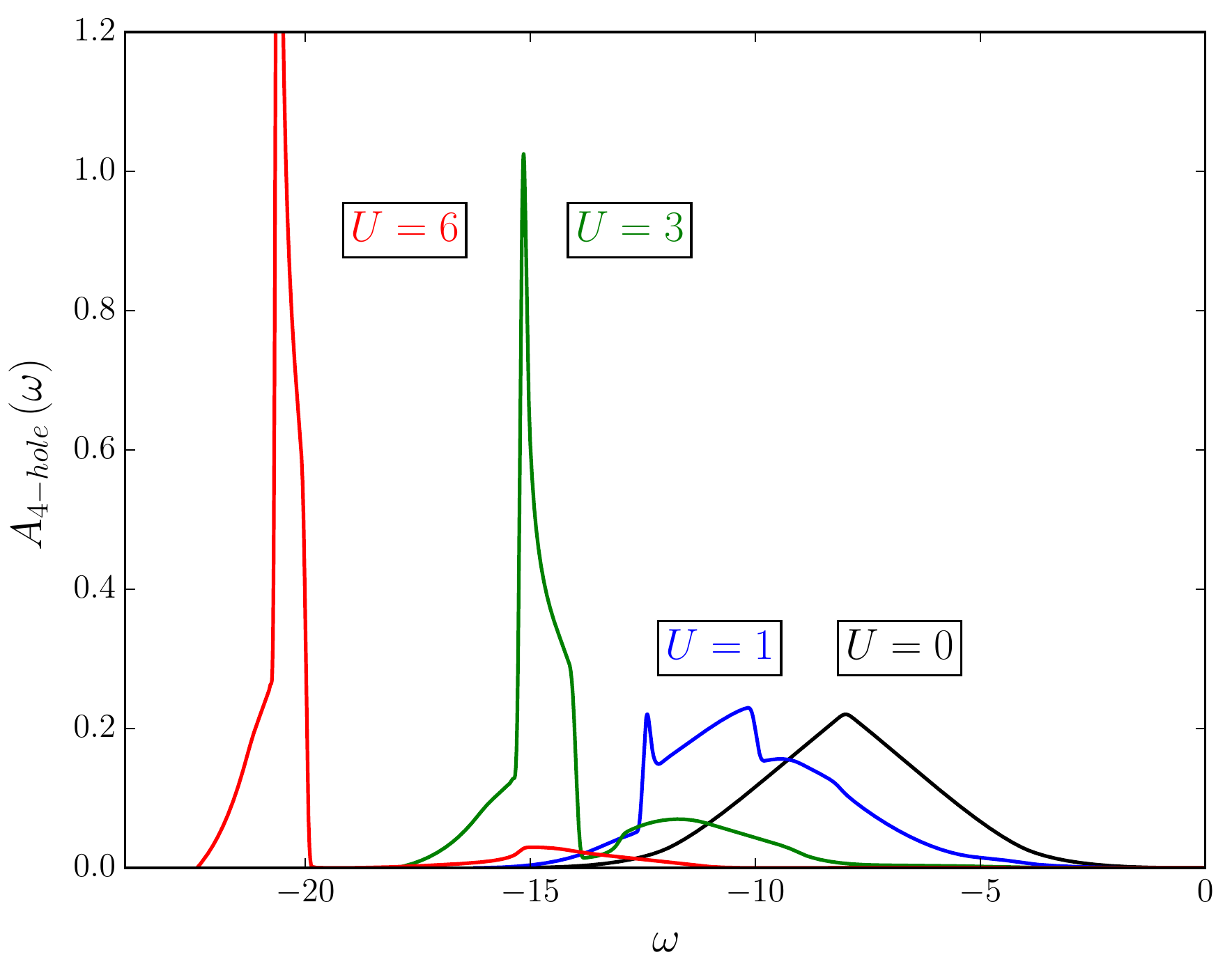}
\caption{
Four-hole excitation spectrum [see Eq.\ (\ref{eq:fourhole})] for $L=100$ sites, different $U$ and $n=2$.
Calculations have been done with resolution $\delta E = 0.05t$ (up to $M=448$ Chebyshev moments) using exact states and periodic boundary conditions.
}
\label{fig:qdr}
\end{figure}

Before addressing this question quantitatively, let us attempt a simple argument favoring an interpretation of ``Q'' as a bound quadruplon opposed to two independently propagating doublons:
Close to half-filling the low-energy states of the system exhibit strong antiferromagnetic correlations. 
Hence, there are two missing magnetic bonds associated with a single hole propagating in this antiferromagnetic background corresponding to an excitation energy $2J$.
Consequently, letting the two neighboring doublons, i.e., the two neighboring holes in the final state (Fig.\ \ref{fig:quad1}b) dissociate and propagate independently costs $4J$ while forming a compound quadruplon requires $3J$ only.
In the strong $J$-limit of the $t$-$J$ model, Eq.\ (\ref{eq:tj}), where the loss of kinetic energy upon quadruplon formation can be neglected, this argument \cite{Gia04} is strict.
For weak $J$, related to the present case of the Hubbard model, it serves as a simple picture for the binding mechanism only. 

A simple inspection of the presence of a binding energy can be done using the position of ``Q'': A two-doublon continuum would be expected at twice the energetic center of the doublon satellite, an additional binding energy would mean a shift from that position. Looking at the closest dataset to half-filling ($n=62/60$ in Fig.\ \ref{fig:nseries}), we do not see such a shift within plotting accuracy, the ``Q'' peak at $\omega \approx 9.8$ coincides with the expected continuum center. Going to higher fillings, however, we find a shift that increases roughly linearly with $n$ (from $\Delta \omega \approx 0.1$ at $n=1.2$ to $\Delta \omega \approx 0.4$ at $n=1.6$) towards lower binding energies, i.e. the doublons seem to bind \emph{attractively} as the filling increases.

For $n \to 2$ the phase space for quadruplon formation shrinks to zero. 
However, we can address the quadruplon formation in the final state and carry out a similar analysis as in Sec.\ \ref{sec:triplon} by calculating the following four-hole spectral function
\begin{eqnarray}
A_{\rm 4-hole}\lr{\omega} 
&=& 
\sum_n \norm{
\matrixel{n,N-4}{c_{i\uparrow}c_{i\downarrow}c_{i+1\uparrow}c_{i+1\downarrow}}{0,N}
} 
\nonumber \\
&\times& 
\delta\lr{\omega + 4 \mu -\lr{E_0^{\lr{N}}-E_n^{\lr{N-4}}}}
\label{eq:fourhole}
\end{eqnarray}
for the completely filled band. 
The result is shown in Fig.\ \ref{fig:qdr}. 

For $U=0$, the spectrum consists of a four-hole continuum at low binding energies; its support is given by $4W=16t$. 
For strong $U$, this continuum has extremely small spectral weight. 
Some weight has been transferred to an extended structure at intermediate excitation energies around $\omega = - 8t - U$ which we identify as the one-doublon-two-hole continuum.
Most of the weight, however, is taken by the correlation satellite located at $\omega \approx - 8t - 2U$ which corresponds to final states with two doublons. 
Notably, its shape does not resemble the self-convolution of a doublon satellite (cf.\ Fig.\ \ref{fig:AES_n=2_Useries}), as one would expect for a doublon-doublon continuum. 

\begin{figure}[t]
\includegraphics[width=\columnwidth]{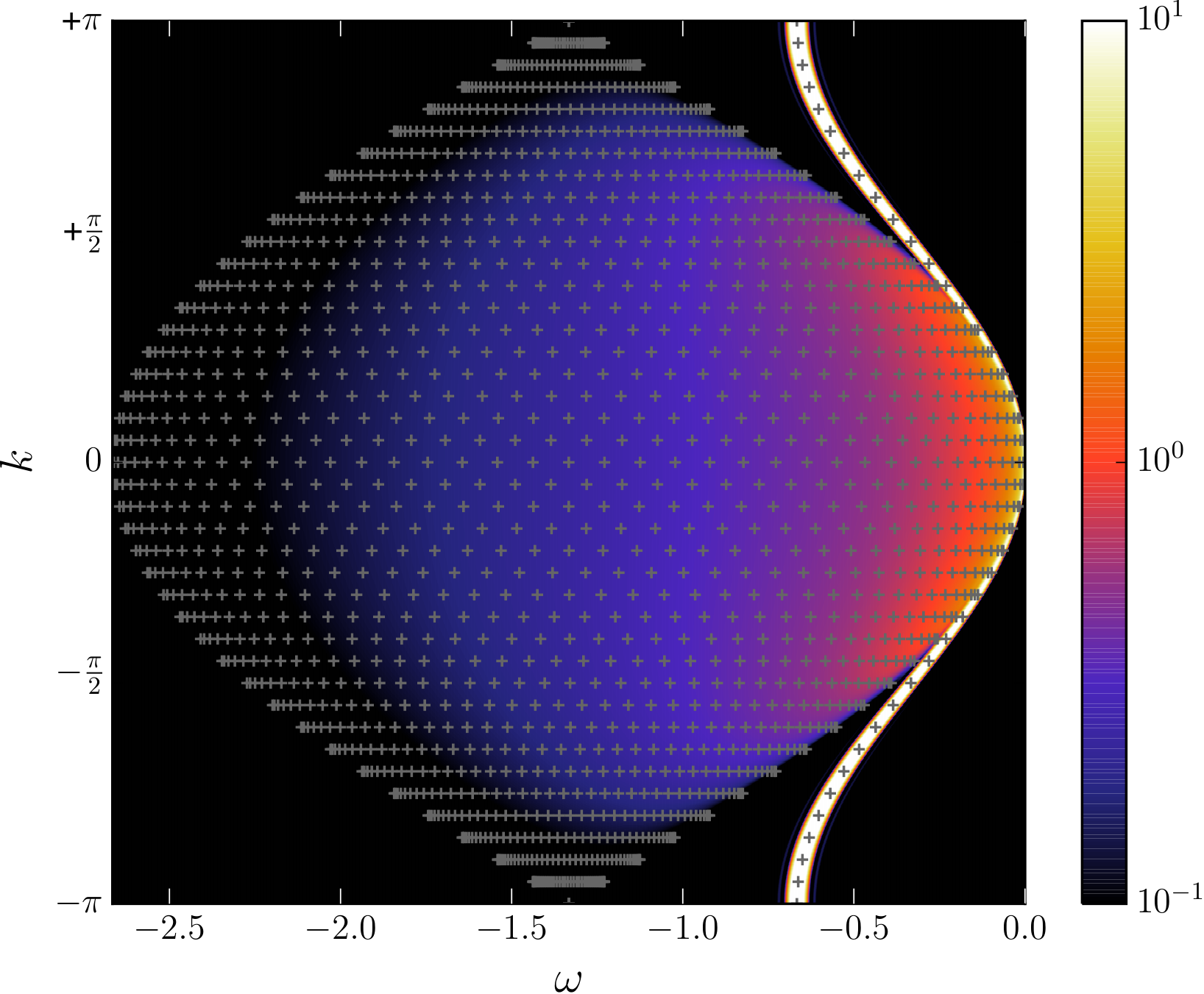}
\caption{
Momentum-resolved two-doublon spectral function [Eq.\ (\ref{eq:2d})] of the effective model [Eq.\ (\ref{eq:Heff})] calculated for $L=1000$ lattice sites at a resolution of $\delta E=0.01t$ using exact states and periodic boundary conditions. 
Calculation for the completely filled system ($N_{d}=L$ bosons) and for coupling constant $J = 2t/3$.
The excitation energies of a smaller system with $L=40$ are shown as crosses for comparison. 
}
\label{fig:qdreff}
\end{figure}

\begin{figure}[b]
\includegraphics[width=\columnwidth]{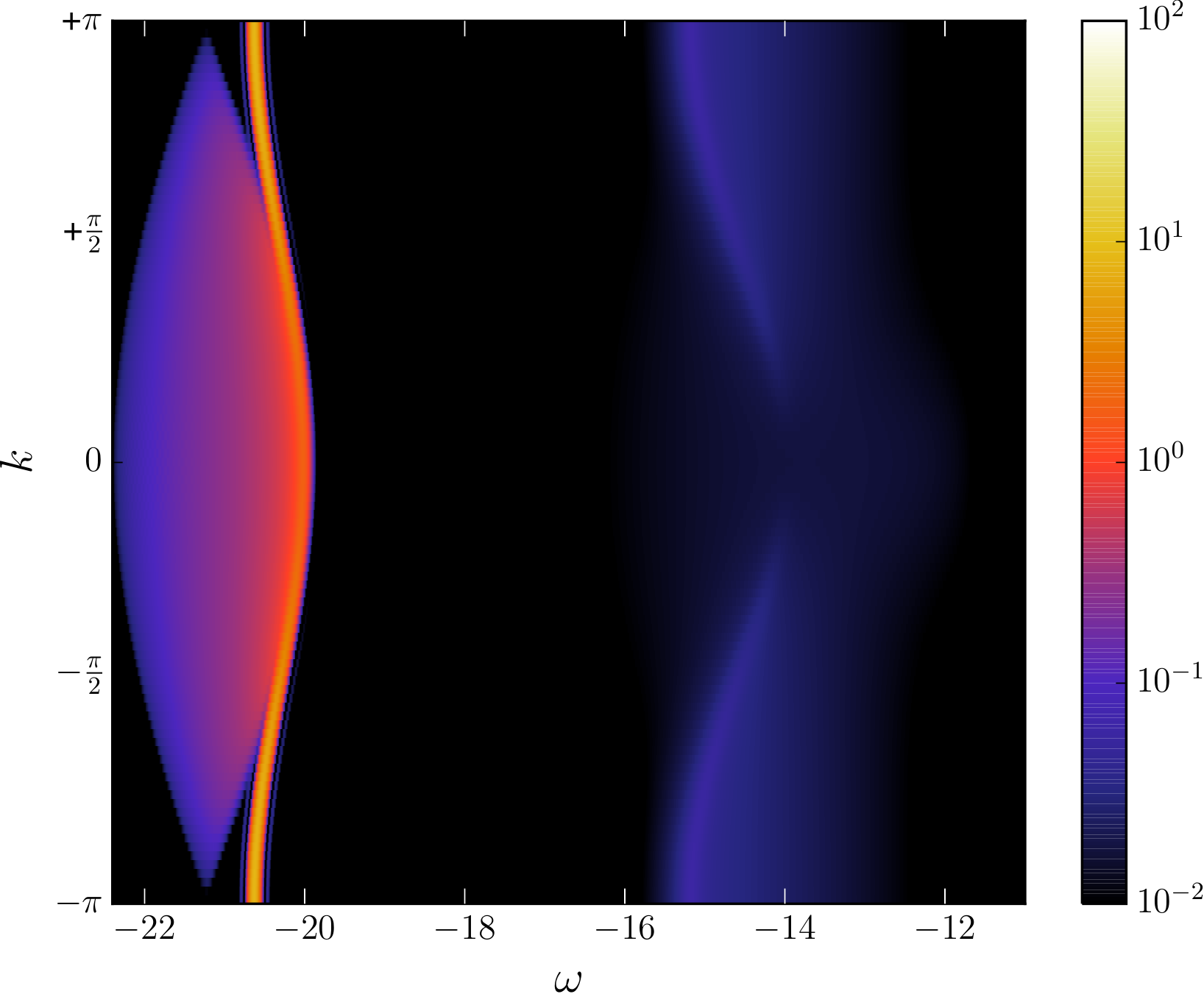}
\caption{
Four-hole spectral function for $U=6t$ and the same parameters as in Fig.\ \ref{fig:qdr}, but $k$-resolved [Eq. (\ref{eq:4h})].
}
\label{fig:QDRk_n=2}
\end{figure}

This is in fact corroborated by the analysis of an effective model \cite{RRBV08,HP12}, valid for strong $U$ in the subspace with a fixed number of doublons, but no singly occupied sites: 
\begin{eqnarray}
  H_{\text{eff}} 
  &=& 
  \frac{J}{2} \sum_{\langle ij \rangle} \lr{d^\dagger_i d_j + \mbox{H.c.}} 
  \nonumber \\
  &+& 
  \lr{J+U} \sum_i n_i^d 
  - 
  J \sum_{\langle ij \rangle} n_i^d n_j^d 
  \: .
\label{eq:Heff}
\end{eqnarray}
The model describes inherently stable bosons with the creation operator $d^\dagger_i = c^{\dagger}_{i\uparrow}c^{\dagger}_{i\downarrow}$ and a hard-core constraint $(d^\dagger_i)^2=0$ enforcing that any site is at most singly occupied with a single boson.
Recall that throughout the paper, a doublon refers to a pair of {\em holes} rather than particles, and hence a doublon at site $i$ corresponds to a hole $n_{i}^{d} = 0$, where $n_i^d = d^\dagger_i d_i$ is the boson occupation-number operator. 
The effective model includes a hopping term with amplitude $J/2 = 2t^2/U$ and a nearest-neighbor density-density interaction of strength $J$, thus twice the hopping amplitude, but only half the bandwidth $2J$.
Note that this effective doublon-doublon interaction is attractive.

Our four-hole problem now becomes a two-hole problem in the effective model, Eq.\ (\ref{eq:Heff}), and can be solved in much the same way as the two-hole problem in the Cini-Sawatzky case, see Ref.\ \onlinecite{SEG94} and Sec.\ \ref{sec:filled}.
Separating the relative from the center-of-mass motion, the eigenvalues can be calculated for rather large systems.
One does indeed observe a split-off resonance for certain values of the wave vector $k$, corresponding to a bound quadruplon. 
However, due to the small interaction strength $J$, the bound state is energetically positioned within the doublon-doublon continuum. 

\begin{widetext}
In order to demonstrate this, rather than showing just the eigenvalues, we calculate the $k$-resolved two-doublon spectral function within the effective model:
\be 
A_{\rm 2-doublon} \lr{\omega,k} 
=
\frac1L
\sum_n \norm{
\matrixel{n,N_{d}-2}{\sum_{i}e^{-ikR_i}d_{i}d_{i+1}}{0,N_{d}}
} 
\delta\lr{\omega -\lr{E_0^{\lr{N_{d}-2}} - E_n^{\lr{N_{d}-2}}}  }
\: .
\label{eq:2d}
\ee
Note that it is for simplicity defined such that the lowest binding energy is $\omega=0$.
Fig.\ \ref{fig:qdreff} shows the result of a calculation for $N_{d}=L=1000$ and $J=2t/3$ corresponding to filling $n=2$ and interaction $U=6t$ in the original model.
The two-doublon spectrum corresponds to the satellite around $\omega = - 8t - 2U - 2J \approx -21.3t$ in Fig.\ \ref{fig:qdr}. 
The support is given by $\Delta \omega = 8t/3 \approx 2.67 t$, though with very little weight on the high-binding-energy side. 
This is the expected value for two independent doublons: $\Delta \omega = 2 \times 4 \times J/2$.

The $k$-resolution uncovers that the continuum actually has very small spectral weight and that the spectrum is in fact dominated by the intense quadruplon peak. 
Around the Brillouin-zone center the quadruplon is a resonance with a finite width, but still its weight dominates the whole energy range. 
For $|k| \gtrsim 3\pi/4$, it is fully split off from the continuum and has zero intrinsic width (the width is due to the finite $\delta E$ only). 
We conclude that the quadruplon excitation represents a stable compound object formed by two bound doublons when its wave vector is close to the zone boundary, while it is a well-defined resonance with finite lifetime around the zone center.

The stability of the quadruplon can be explained by a phase-space argument:
For a state from the two-doublon continuum, the hard-core constraint is not effective in the thermodynamic limit.
Hence, the total energy of two doublons with wave vectors $p$ and $k-p$ is obtained as $J \cos(p) + J \cos(k-p)$. 
Right at $k=\pm \pi$, this adds up to zero, so that a bound quadruplon of arbitrary energy $\omega \ne 0$ and wave vector $k=\pm \pi$ cannot decay.
Note that the quadruplon branches split off from the continuum on the low-binding-energy side, as the quadruplon is bound by an {\em attractive} interaction of strength $J$, see Eq.\ (\ref{eq:Heff}). 
Indeed, the quadruplon binding energy at $k=\pm \pi$ is given by $J = 2t^{2}/3U \approx 0.67 t$.\cite{SEG94}

The same physics is found when considering the $k$-resolved four-hole spectral function in the full model:
\be
A_{\rm 4-hole}\lr{\omega,k} 
= 
\frac1L
\sum_n 
\norm{
\matrixel{n,N-4}{\sum_i e^{-ikR_i}c_{i\uparrow}c_{i\downarrow}c_{i+1\uparrow}c_{i+1\downarrow}}{0,N} 
}
\delta\lr{\omega + 4 \mu -\lr{E_0^{\lr{N}} - E_n^{\lr{N-4}}}}
\: .
\label{eq:4h}
\ee
This is shown in Fig.\ \ref{fig:QDRk_n=2} for $U=6t$. 
Due to the now allowed singly occupied sites and a finite $U$, the one-doublon-two-hole continuum around $\omega \approx \omega = - 8t - U$ acquires some spectral weight. 
This corresponds to final states where one of the doublons has decayed.
Most of the spectral weight, however, is taken by the quadruplon resonance or bound state at higher binding energies.
For the quadruplon binding energy at $k=\pm \pi$, we obtain about $0.58t$ due to the moderate strength of $U=6t$.
\end{widetext}

\begin{figure}[t]
\includegraphics[width=\columnwidth]{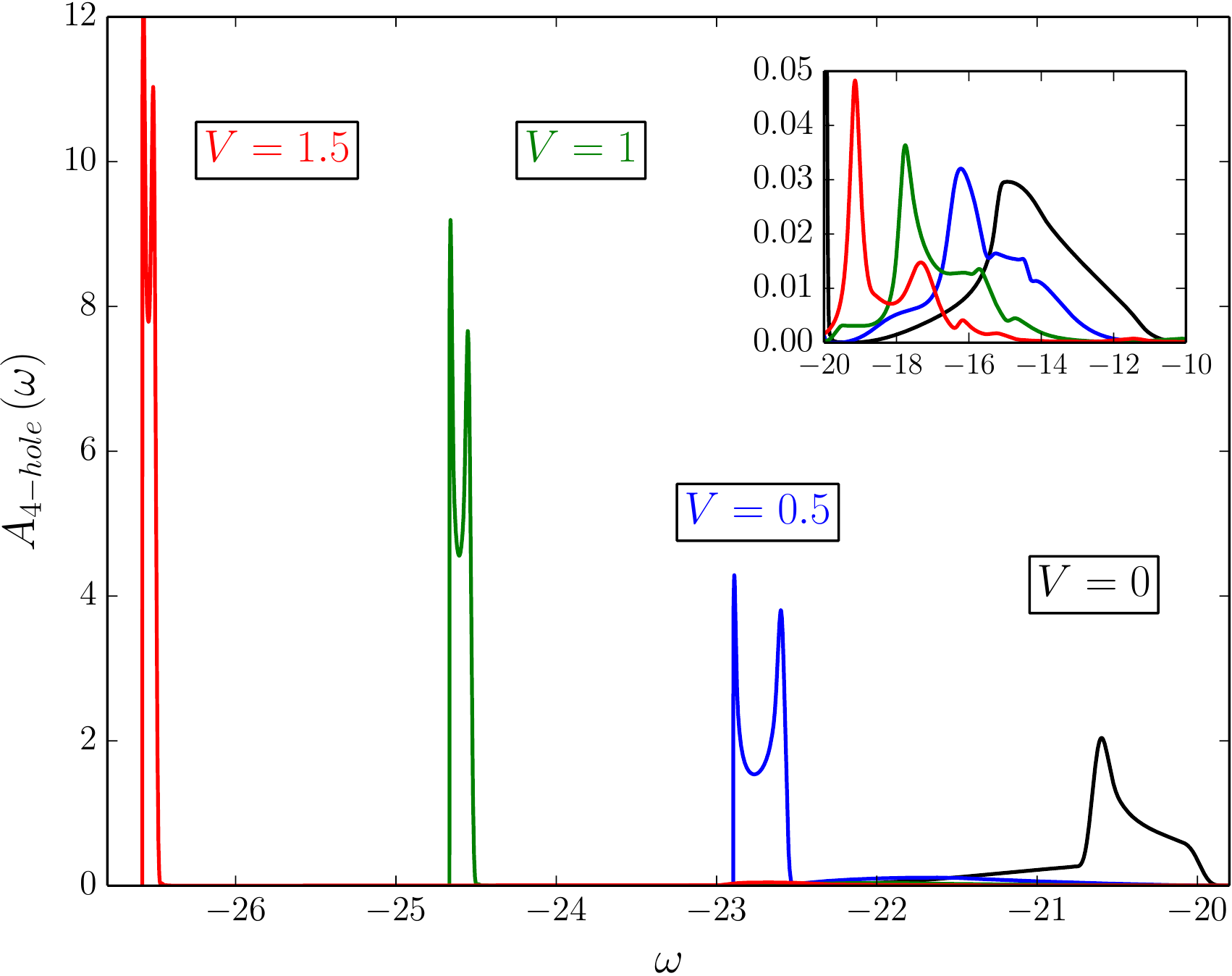}
\caption{Four-hole spectral function for $L=100$ sites, a resolution $\delta E = 0.05t$ (up to $M=531$ Chebyshev moments) and a completely filled band ($n=2$) for several values of $V$ ($U=6t$) using exact states and periodic boundary conditions. The inset shows a close-up view of the continua.}
\label{fig:QDR_n=2_Vseries}
\end{figure}

At this point, we briefly note that the quadruplon is stabilized in the entire Brillouin zone by a sufficiently strong nearest-neighbor Coulomb interaction $V>0$, very similar to the case of the triplon excitation discussed in Sec.\ \ref{sec:triplon}. 
Fig.\ \ref{fig:QDR_n=2_Vseries} shows the (local) four-hole spectrum for $U=6t$ and different $V$. 
Here, the quadruplon satellite, with a shape approaching the shape of the free density of states for strong $V$, is fully split off from the (very weak) two-doublon continuum and shows up on its high-binding-energy side since the interaction is repulsive.
With increasing $V$, the quadruplon peak shifts with the additional binding energy $4V$.

Next, we continue to tackle the question of whether there is a stable bound quadruplon in the {\em two-hole} spectrum. 
First, we note that the structure ``Q'' in $A_{\rm 2-hole}(\omega)$ (see Fig.\ \ref{fig:nseries}) continuously connects to the quadruplon peak analyzed with $A_{\rm 4-hole}(\omega)$. 
This is demonstrated in Appendix \ref{sec:four} where the (local) four-hole spectrum is presented in the entire filling range $1\le n \le 2$.

Hence, the question to be answered is the following: 
Do the two doublons in the final states contributing to ``Q'' for fillings $n<2$ bind and form a compound object, i.e., is there an enhanced probability to find the doublons at neighboring sites?
Above we have already given an argument relying on the energy shift of ``Q'' with respect to the expected position of the two-doublon continuum. 
However, a more direct analysis should be possible as the numerical approach is based on the many-body wave function which, in principle, contains all information on the system.
One should therefore be able to extract specific information on the final states within a given frequency range.
An implementation of this idea within the Chebyshev expansion method requires a new technique as described below.

\subsection{Filter-operator technique}
\label{sec:filter}

To extract specific information on the states contributing to the two-hole spectrum at a given frequency $\omega$, we proceed in the following way: 
First, we define a state $\ket{\Psi(\omega)}$ as the sum over the complete set of energy eigenstates weighted by the matrix element of the transition operator:
\be
  \ket{\Psi(\omega)} 
  = 
  \sum_m \matrixel{m}{c_{i\uparrow}c_{i\downarrow}}{0}
  \, 
  \delta\lr{\omega-\lr{E_0-E_m}} \ket{m} 
  \: .
\ee
$\ket{\Psi(\omega)}$ is constructed to represent a typical final state contributing to the spectrum at frequency $\omega$. 
Note that it is coincidentally identical to the ``correction vector'', 
\be
  \ket{\Psi(\omega)} 
  = 
  - \frac{1}{\pi} \text{Im} 
  \frac{1}{\omega+i0^+-(E_0-H)} 
  c_{i\uparrow}c_{i\downarrow} \ket{0} 
  \: ,
\ee
which is employed within dynamical DMRG\cite{Jec02} to calculate the spectral function via $A_{\rm 2-hole}(\omega) = \langle 0 | c_{i\downarrow}^{\dagger} c_{i\uparrow}^{\dagger} | \Psi(\omega)\rangle$.

\begin{widetext}
Here, we employ it in a different way, choosing a Hermitian (but otherwise in principle arbitrary) ``filter operator'' $F$ and calculate the following expectation value:
\be
\matrixel{\Psi(\omega)}{F}{\Psi(\omega)} 
= 
\sum_{mn} 
\matrixel{0}{c^{\dagger}_{i\downarrow}c^{\dagger}_{i\uparrow}}{m} 
\matrixel{m}{F}{n} 
\matrixel{n}{c_{i\uparrow}c_{i\downarrow}}{0}
\,
\delta\lr{\omega-\lr{E_0-E_m}} 
\delta\lr{\omega-\lr{E_0-E_n}}
\: .
\label{eq:AESfilter1}
\ee
This constitutes a ``filtered'' two-hole spectral function which we call $A_{\rm 2-hole}\left[F\right]\lr{\omega}$. The additional weight factors $\matrixel{m}{F}{n}$ give us some characterization of the eigenstates involved, depending on the choice of $F$. 
The normal spectrum is obviously recovered for $F=1$. 
This quantity is numerically accessible as the diagonal part ($\omega=x=y$) of the two-dimensional Chebyshev expansion defined by
\be
A_{\rm 2-hole}\left[F\right]\lr{x,y} 
= 
\sum_{mn} 
\matrixel{0}{c^{\dagger}_{i\downarrow}c^{\dagger}_{i\uparrow}}{n} 
\matrixel{m}{F}{n} 
\matrixel{n}{c_{i\uparrow}c_{i\downarrow}}{0} 
\, 
\delta\lr{x-\lr{x_0-x_n}} 
\delta\lr{y-\lr{x_0-x_m}}
\: ,
\label{eq:AESfilter2}
\ee
where the moments are given by
\be
\mu_{mn} = \sum_{mn} \matrixel{0}{c^{\dagger}_{i\downarrow} c^{\dagger}_{i\uparrow} ~ T_m(\widetilde{H}) ~ F ~ T_n(\widetilde{H}) ~ c_{i\uparrow}c_{i\downarrow}}{0} 
\ee
in terms of the Chebyshev polynomials of the rescaled Hamiltonian, see Sec.\ \ref{sec:mod}.
\end{widetext}

To convince ourselves that this method works, we first consider the well-understood limit of the completely filled band $n=2$. We define an operator $h_i$ which projects out states with an empty site $i$ as
\be
  h_i = \lr{1-n_{i\uparrow}} \lr{1-n_{i\downarrow}} \: ,
\ee
and an operator $s_i$, projecting out states with single occupancy at $i$:
\be
s_i = \lr{1-n_{i\uparrow}n_{i\downarrow}} \lr{1-h_i} = n_{i\uparrow} + n_{i\downarrow} - 2 n_{i\uparrow} n_{i\downarrow}
\: .
\ee
With this we can use $F=\frac{1}{L} \sum_i h_i$ to filter out states characterized by doublons, and $F=\frac{1}{L} \sum_i s_i$ to filter out states characterized by the absence of doublons, but the presence of single holes. 

\begin{figure}[t]
\includegraphics[width=\columnwidth]{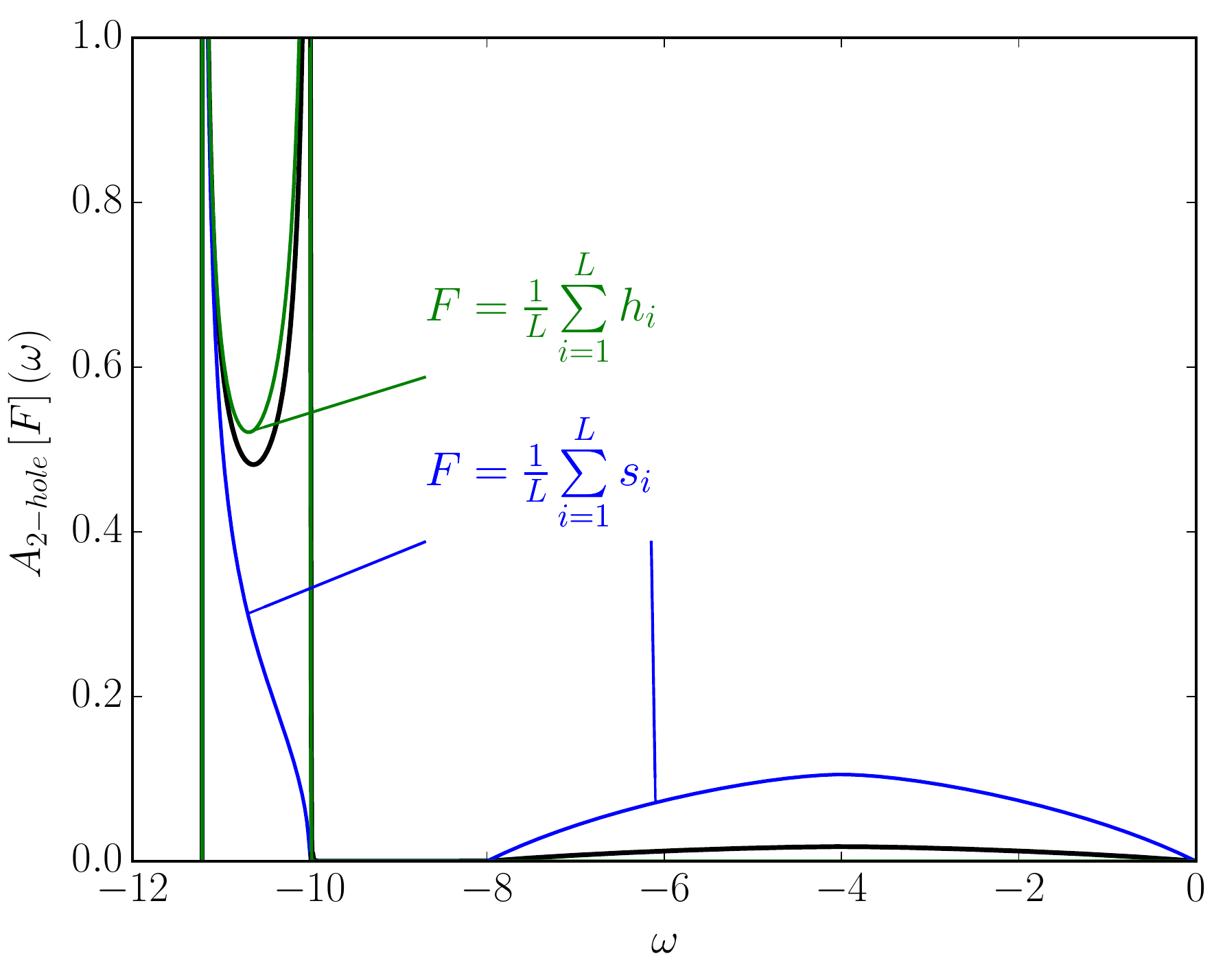}
\caption{
Two-hole spectral function for $n=2$ (thick black line) and two-hole spectra (normalized to area) using different filter operators (colored lines), calculated with exact states for $U=6t$, $L=1000$, $\delta E=0.01t$ ($M=1121$).
}
\label{fig:filter_n=2}
\end{figure}

\begin{figure}[t]
\includegraphics[width=\columnwidth]{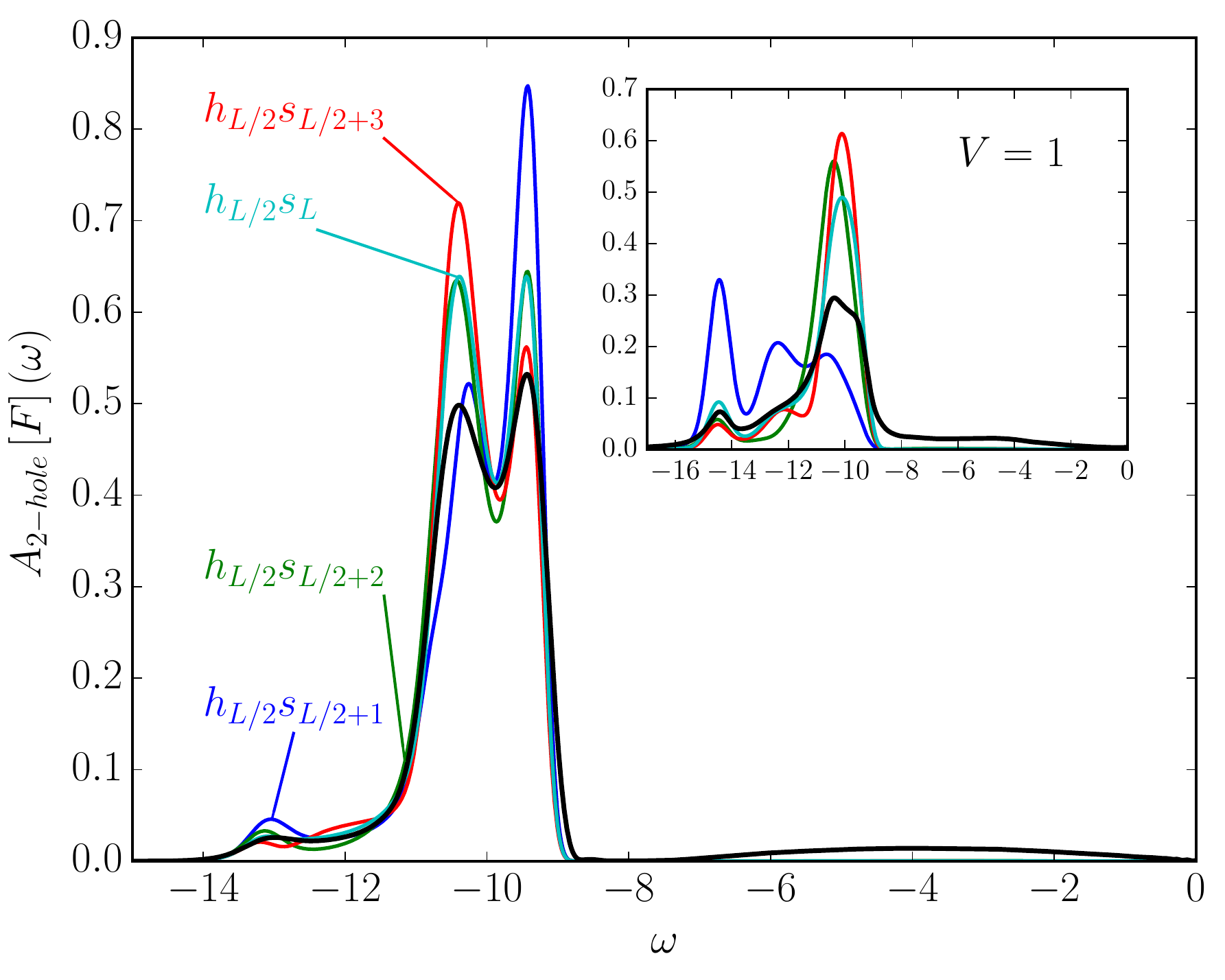}
\caption{
Two-hole spectral function for $n=1.8$ (thick black line) and two-hole spectra (normalized to area) using different filter operators (colored lines, see text), calculated with DMRG for $U=6t$, $L=60$, $\delta E=0.2t$ ($M=380$).
Inset: results for finite nearest-neighbor interaction $V=1$, at $\delta E=0.4t$ ($M=239$).
}
\label{fig:filter_TRI}
\end{figure}

\begin{figure}[b]
\includegraphics[width=\columnwidth]{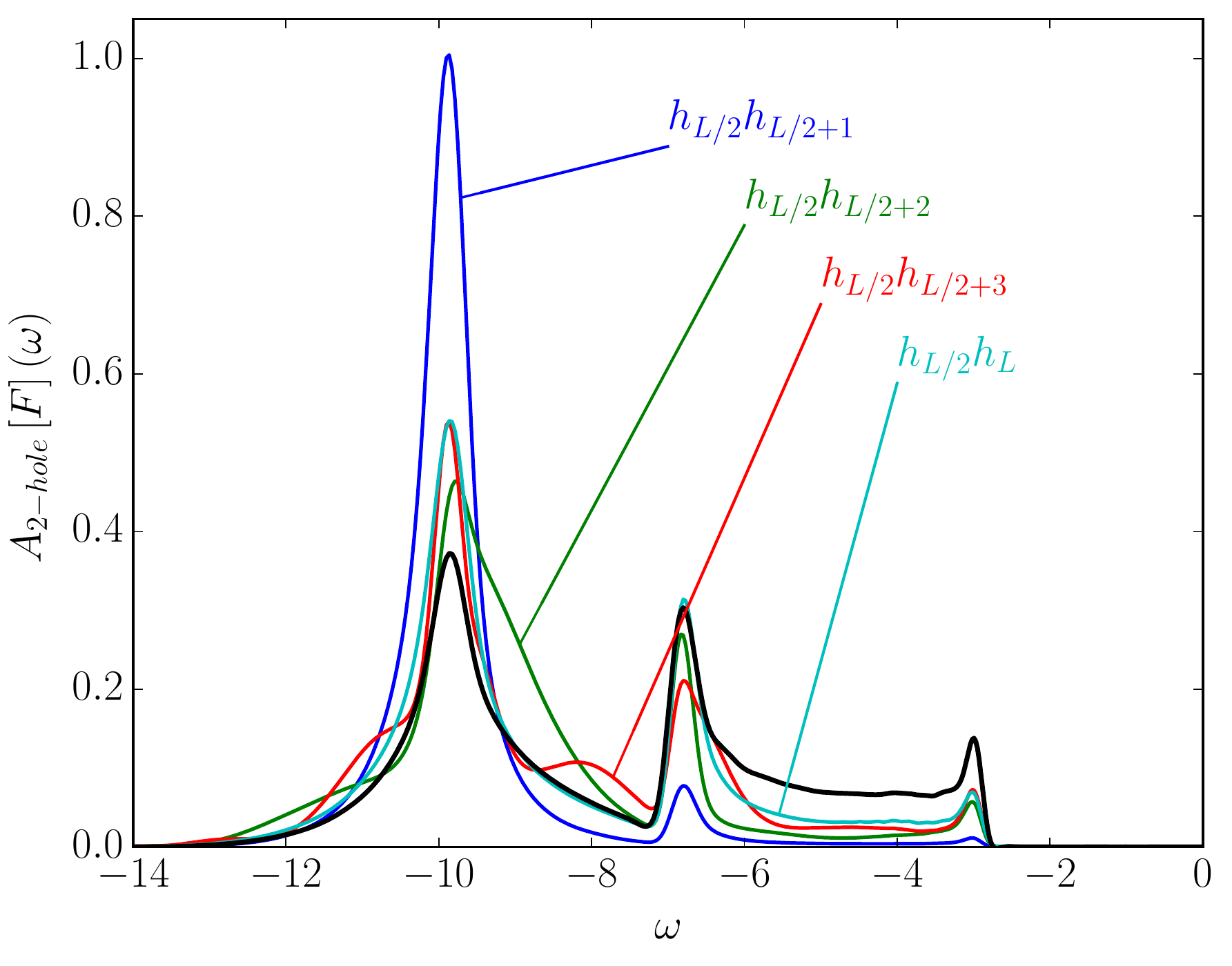}
\caption{
Two-hole spectral function for $n=62/60 \approx 1.03$ (thick black line) and two-hole spectra (normalized to area) using different filter operators (colored lines, see text), calculated for $U=6t$, $L=60$, $\delta E=0.3t$ ($M=766$ Chebyshev moments) using DMRG.
}
\label{fig:filter_QDR}
\end{figure}

Results for $U=6t$ are shown in Fig.\ \ref{fig:filter_n=2}. 
We do indeed observe that the spectral function obtained with the former filter has support at the doublon satellite only, while the support of the latter coincides with spectral range of the band-like part and of the satellite. 
That the spectral function filtering out final states with singly occupied sites has finite weight at the doublon satellite reflects that doublon propagation in fact involves intermediate configurations with singly occupied sites.

Next we test the triplon case. 
Recall that the two-hole spectral function is computed at the central site of the chain, i.e., we take $i=L/2$ for the transition operator $c_{i\uparrow}c_{i\downarrow}$. 
To look for triplons, we can thus choose local filters and take $F = h_{L/2} s_{L/2+\delta}$ at different distances $\delta$. 
An enhanced weight at $\delta=1$ would indicate that the doublon and the additional hole bind and form a stable triplon. 

The result of calculations for $n=1.8$ is shown in Fig.\ \ref{fig:filter_TRI}. 
We do indeed see that the case $\delta=1$ does not look any more special than $\delta=2$, $\delta=3$ or $\delta=L/2$.
All choices roughly reproduce the shape of the two-hole spectral function and are barely distinguishable in the range of the peak at $\omega \approx -13.1t$, earlier identified as the triplon structure.
This corroborates our previous result that doublon and hole do not bind.
On the other hand, a stable triplon is formed for a nearest-neighbor interaction $V=1$. 
This is indeed supported by the filter-operator technique: 
For $V=1$ (see inset of Fig.\ \ref{fig:filter_TRI}) the case $\delta=1$ is in fact very different as compared to $\delta \ne 1$: 
The weight of the filtered spectrum is strongly enhanced at frequencies around $\omega=-14.4t$, where the peak of the bound triplon is located (cf.\ Fig.\ \ref{fig:aesv}).

Finally, to investigate the quadruplon, we consider the filling $n=62/60$ right above half-filling.
The quadruplon peak is located at $\omega \approx -9.8t$ below the doublon structure with barycenter at $\omega\approx - 6t$, see Fig.\ \ref{fig:nseries}.
We have computed filtered spectra using $F = h_{L/2} h_{L/2+\delta}$. 

Results are shown in Fig.\ \ref{fig:filter_QDR}. 
In this case, a filter with $\delta=1$ produces a strong signal at the quadruplon position even for $V=0$, while the spectra for larger distances $\delta \ne 1$ are again close to each other and roughly coincide with the original spectral function. 
This is a clear indication that there is a bound quadruplon, the contribution of which dominates the two-hole spectrum $A_{\rm 2-hole}(\omega)$, as compared to the doublon-doublon continuum. 

\subsection{Infinite doublon lifetime for $k=\pi$}
\label{sec:res:kdependence}

Let us return to the doublon. 
As has been discussed in Sec.\ \ref{sec:doublon}, a stable excitation with infinite lifetime is found in the simple $n=2$ limit and explained as a repulsively bound pair of holes.
This interpretation breaks down for fillings close to half-filling, where the two-hole excitation propagates through a background of singly occupied sites with antiferromagnetic correlations. 
We are thus left with the question whether there is a stable doublon for fillings $n<2$, and close to half-filling in particular. 

The $U$-dependence of the doublon lifetime has been addressed in a number of previous theoretical as well as experimental studies, \cite{SGJ+10,LP13,CGK12,PSAF07,HP12,RRBV08,HPS11} starting from the concept of repulsive binding. 
Here, we would like to emphasize that the $k$-resolution of the local two-hole excitation adds an important new view to this problem.
For the case of periodic boundary conditions, the $k$-resolved two-hole spectral function is defined as:
\begin{eqnarray}
A_{\rm 2-hole}\lr{\omega,k} 
& = &
\frac1L
\sum_n 
\norm{
\matrixel{n,N-2}{\sum_{i} e^{-ikR_{i}} c_{i\uparrow}c_{i\downarrow}}{0,N}
} 
\nonumber \\
&\times&
\delta \lr{\omega + 2 \mu - \lr{E_0^{\lr{N}} - E_n^{\lr{N-2}} }}
\: .
\label{eq:spectrumk}
\end{eqnarray}
Exploiting translational symmetry, it is straightforward to show that $k$-summation yields 
\be
A_{\rm 2-hole}\lr{\omega} 
=
\frac1L \sum_{k}
A_{\rm 2-hole}\lr{\omega,k} 
\label{eq:ksum}
\: .
\ee
Hence, with $A_{\rm 2-hole}\lr{\omega,k}$, one actually decomposes the local two-hole excitation into a coherent linear superposition of $k$-dependent two-hole excitations:
\be
c_{i\uparrow}c_{i\downarrow} \ket{0,N}
=
\sum_{k} d_{k} \ket{0,N}
\ee
where $d_{k} = L^{-1} \sum_{i} e^{-ikR_{i}} c_{i\uparrow} c_{i\downarrow}
= L^{-1} \sum_{p} c_{p\uparrow} c_{k-p\downarrow}$ with 
$c_{k\sigma} = L^{-1/2} \sum_{i} e^{-ikR_{i}} c_{i\sigma}$.
The important observation is that the state $d_{k} \ket{0,N}$ has a strongly $k$-dependent lifetime. 
At the zone boundary $k=\pi$, the lifetime is even infinite. 

\begin{figure}[t]\
\includegraphics[width=0.9\columnwidth]{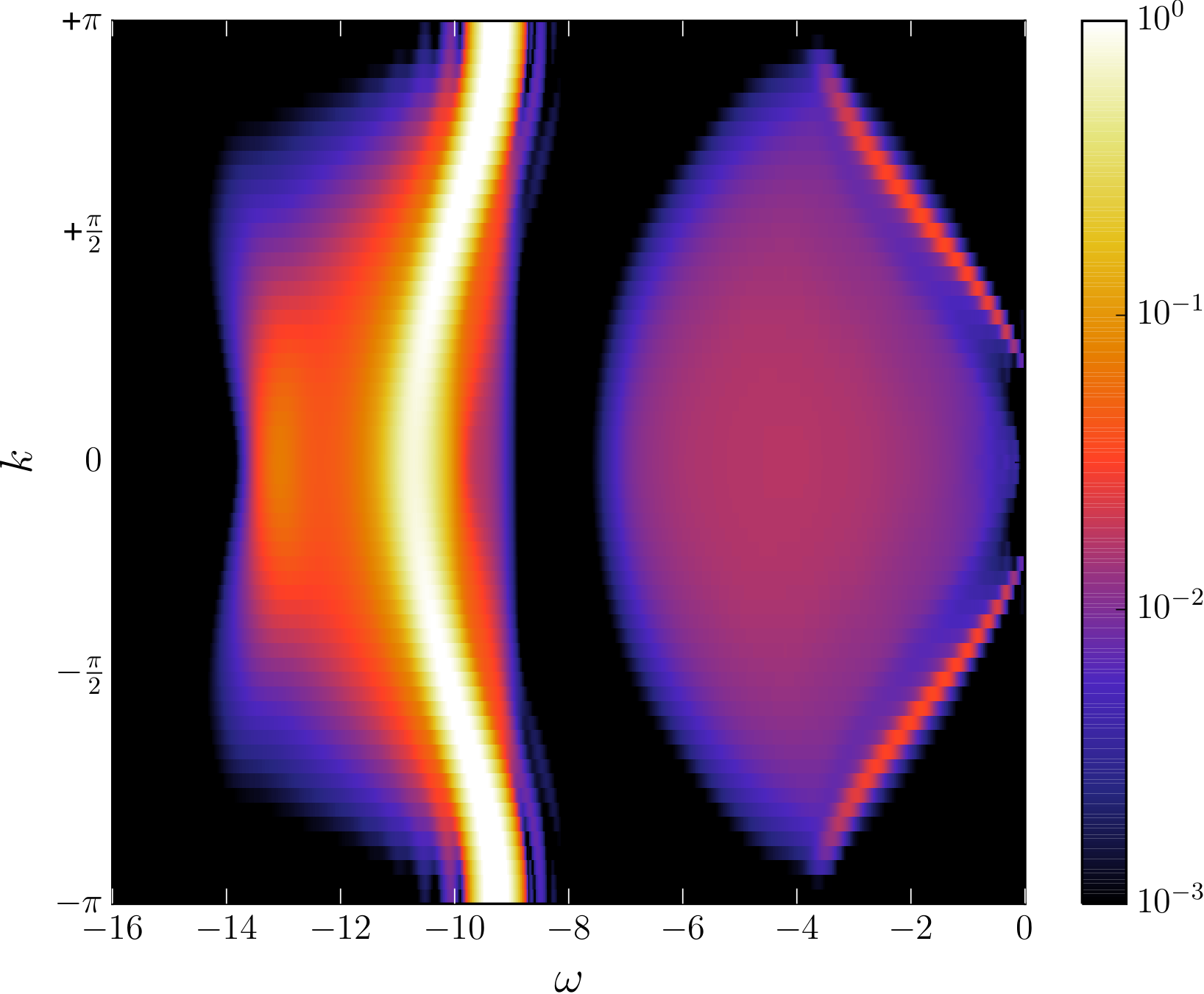}
\caption{
Momentum-resolved two-hole spectral function for $n=1.8$ and $U=6t$.
DMRG calculations for $L=60$ sites, $\delta E = 0.2t$ ($M=380$ Chebyshev moments).
Note that the spectral weight is plotted on a logarithmic scale. 
}
\label{fig:AESk_n=1.8}
\end{figure}

This can be verified numerically and understood analytically.
Let us first discuss the numerical results. 
Eq.\ (\ref{eq:spectrumk}) can be used to approximate the two-hole spectrum with a calculation for a system with open boundaries. 
In this case, the $k$-summation of $A_{\rm 2-hole}(\omega,k)$ [Eq.\ (\ref{eq:ksum})] yields the local spectrum, but averaged over all sites, rather than the local spectral function at the central site $i$, as defined in Eq.\ (\ref{eq:A2hole}). 
For a system with size $L=60$, however, the approximation is already excellent. 
Note that this can also be thought of as carrying out a Fourier transform with two momenta, $A_{\rm 2-hole}(\omega,k,k')$, but then only looking at the diagonal component with $k=k'$: $A_{\rm 2-hole}(\omega,k) \equiv A_{\rm 2-hole}(\omega,k,k)$.

Fig.\ \ref{fig:AESk_n=1.8} shows the $k$-resolved spectrum for $n=1.8$ and $U=6t$. 
The band-like part extends from $\omega=0$ to $\omega\approx - 8t$ with most of the spectral weight at low binding energies and a ridge-like structure around $k=0$. 
Its overall spectral weight, however, is very small compared to the weight of the doublon satellite, which is centered around $\omega \approx -10t$.
Interestingly, one finds that the doublon satellite is not broadened at all for $k=\pm\pi$. 
In fact, the structure at $k=\pm\pi$ should be interpreted as a $\delta$-peak at frequency $\omega = \omega_{0}(\pi) \approx -9.25t$. 
Note that on the logarithmic scale, there are seemingly sidebands visible close to $\omega_{0}(\pi)$ and $k=\pm \pi$. 
These are, however, simply the numerical artifact of Gibbs oscillations.
As shown below, the position of the $\delta$-peak is precisely given by $\omega_{0}(\pi) = U - 2\mu \approx -9.25t$, where the value of $\mu$ is obtained from the ground-state calculation [Eq.\ (\ref{eq:mu})].
When $|k|<\pi$, the satellite acquires some finite width, i.e., the doublon transmutes from a stable bound object into a resonance. 
Its lifetime is shortest around $k=0$.
This is opposed to the simple $n=2$ limit, where the doublon is stable in the entire Brillouin zone.

\begin{figure}[t]
\includegraphics[width=0.9\columnwidth]{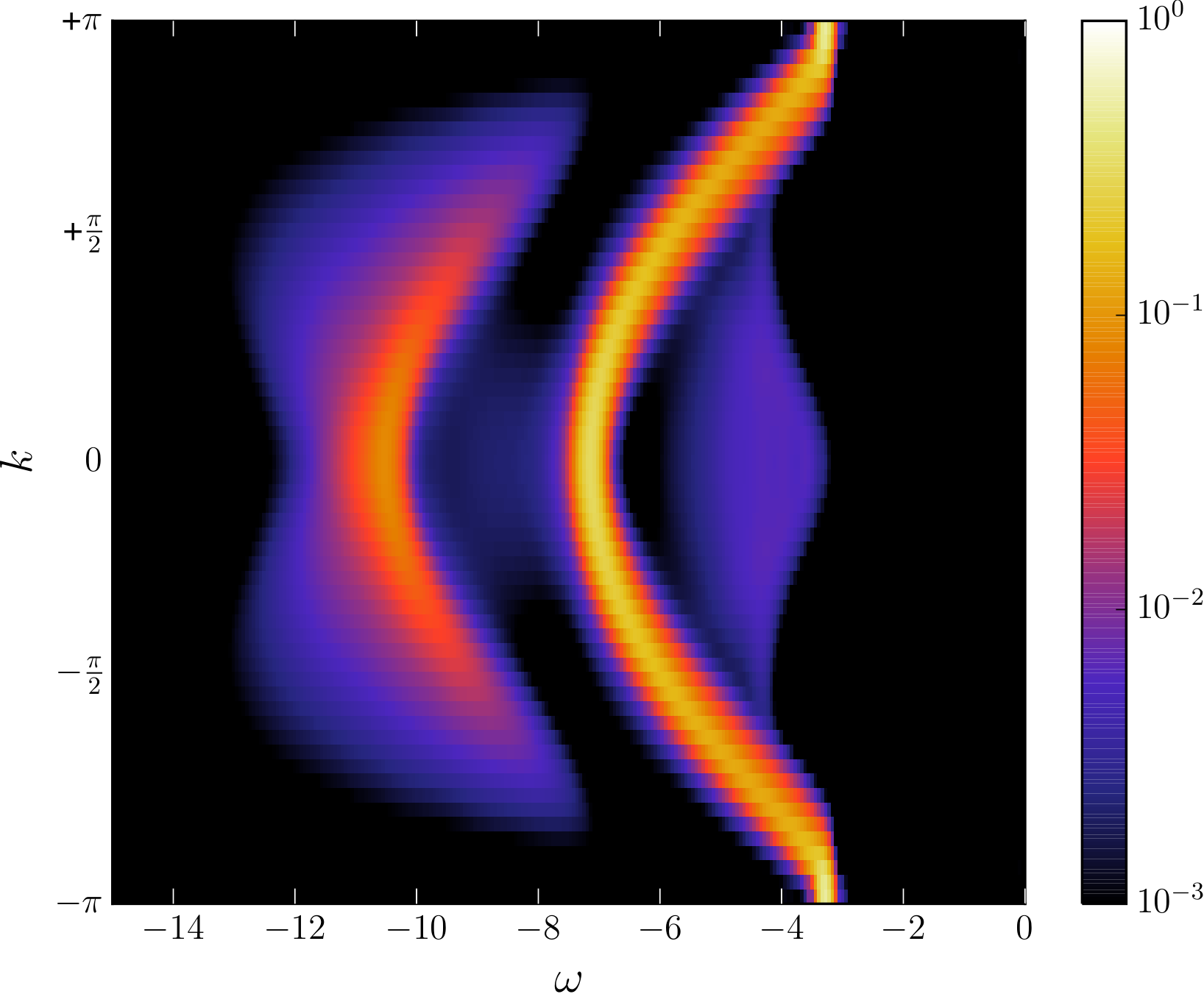}
\caption{
The same as Fig.\ \ref{fig:AESk_n=1.8}, but for $n=1.1$ ($M=1118$).
}
\label{fig:AESk_n=1.1}
\end{figure}

The situation at a filling $n=1.1$ is very similar. 
Results are shown Fig.\ \ref{fig:AESk_n=1.1}.
Here, the $k$-resolved two-hole spectrum consists of an intense doublon satellite (centered around $\omega \approx -5t$) and the quadruplon structure (around $\omega \approx -10t$), as well as a band-like part with non-zero, but extremely low weight close to $\omega = 0$ which is not visible in the figure.
Referring to the results obtained by means of the filter-operator technique in Sec.\ \ref{sec:filter}, we interpret the quadruplon structure as an object composed of two doublons but with a finite lifetime, i.e., a resonance. 
Most of its weight is located close to $k=0$. 
It is tempting to assume that the quadruplon is entirely stable at $k=\pi$, similar to the four-hole spectrum case (Fig.\ \ref{fig:QDRk_n=2}), but the whole spectral weight at the zone boundary is instead taken by the doublon satellite. We have already seen a similar effect in the case of the triplon: In the three-hole spectral function for $n=2$ (Fig.\ \ref{fig:tri2v}), it clearly has a different dispersion from the two-hole spectral function for $n<2$ (the inset of Fig.\ \ref{fig:aesv}), while the energetic positions follow the same rules.

Turning to the doublon in Fig.\ \ref{fig:AESk_n=1.1}, we observe that it has infinite lifetime at $k=\pm \pi$ again, but now also displays a distinct ``shadow'' around $k=0$. 
For $n=1.1$ we find $2\mu=9.25$, and thus expect the doublon at energy  $\omega_{0}(\pi) = U-2\mu \approx -3.25t$. 
This perfectly fits with the position that can be read off from Fig.\ \ref{fig:AESk_n=1.1}.

The stability of the doublon at the zone edges can be understood analytically by referring to the ``hidden'' charge-SU(2) symmetry of the Hubbard model. \cite{Yan89,Zha90}
We define 
\be
  {\eta} = \sum_i \lr{-1}^i c_{i\uparrow} c_{i\downarrow} \: .
\label{eq:eta}  
\ee
For the model considered here, and in general for a Hubbard model with $V=0$ on a bipartite lattice in $D$ dimensions, one easily verifies that $\eta$ is an eigenoperator of $H$ and of the total particle-number operator $\hat{N}$: 
\be
  [{\eta}, H] =  U \eta \; , \quad
  [{\eta}, \hat{N}] =  2 \eta \: . 
\label{eq:com}
\ee
$\eta$ is related to the total isospin $\ff T$, the components of which are given by $T_{x} = (\eta + \eta^{\dagger})/2$, $T_{y} = (\eta - \eta^{\dagger})/(2i)$ and $T_{z} = (L - \hat{N})/2$, and represent the generators of a global charge-SU(2) symmetry of the model at half-filling: $[\ff T, H-\mu \hat{N}] = 0$ for $\mu = U/2$.
Starting from the total spin $\ff S$, which generates the standard spin-SU(2) group, the isospin is obtained by a spin-asymmetric and staggered particle-hole (Shiba) transformation $U_{\rm sh}$ with $U_{\rm sh}^{\dagger} c_{i\uparrow} U_{\rm sh} = (-1)^{i} c^{\dagger}_{i\uparrow}$.
We have $U_{\rm sh}^\dagger \ff S U_{\rm sh} = \ff T$.

At half-filling ($N=L$) the ground state $\ket{0,N}$ is a non-degenerate total spin singlet. 
This implies immediately that $\ket{0,N}$ is a total isospin singlet as well.
Hence, we have $\eta \ket{0,N=L} = 0$.
Since $d_{k=\pi} = \eta / L$ for $k=\pi$, the transition operator in the Lehmann representation of the $k$-resolved two-hole spectral function, Eq.\ (\ref{eq:spectrumk}), is just proportional to $\eta$. 
For the case of half-filling this implies
\be
  A_{\rm 2-hole}(\omega, k=\pi) = 0 
\label{eq:api}
\ee
for all excitation energies $\omega$. 
This is consistent with the DMRG calculations which predict the doublon satellite to vanish for $n=1$, even in the entire Brillouin zone (see Fig.\ \ref{fig:nseries}). 

Off half-filling, the chemical potential $\mu \ne U/2$ explicitly breaks the charge-SU(2) symmetry and produces a massive collective mode showing up as a single $\delta$-peak in the two-hole spectrum at wave vector $k=\pi$ (the same holds for $k=-\pi$).
This can be inferred from Eq.\ (\ref{eq:com}).
Namely, it is easy to see that, for arbitrary $N$, the state $\eta \ket{0,N}$ is an {\em exact} eigenstate of $H$ with energy $E_{0}^{(N)} - U$.
Inserting this into Eq.\ (\ref{eq:spectrumk}), yields
\be
  A_{\rm 2-hole}(\omega, k=\pi) = \alpha \:  \delta(\omega + 2\mu - U) \: .
\label{eq:delta}
\ee
This explains the observed position $\omega_{0}(\pi) = U - 2 \mu$.

Integrating the $k$-resolved spectrum, Eq.\ (\ref{eq:spectrumk}), over $\omega$, yields the total weight at a given $k$:
\be
  \alpha(k) = \frac1L \sum_{ij} 
  e^{-ik(R_{i} - R_{j})}
  \langle 0,N | 
  c^{\dagger}_{j\downarrow} c^{\dagger}_{j\uparrow} 
  c_{i\uparrow} c_{i\downarrow} 
  | 0,N \rangle \: .
\label{eq:alpha}  
\ee
This gives $\sum_{k} \alpha(k) = \sum_{i} \langle n_{i\uparrow} n_{i \downarrow} \rangle$, consistent with the global sum rule Eq.\ (\ref{eq:sumrule}). 
At $k=\pi$, as expressed by Eq.\ (\ref{eq:delta}), the whole spectral weight condenses to $\alpha(\pi)$ at frequency $\omega = U - 2\mu$.
The weight of the mode at the zone boundary is then given by $\alpha = L^{-1} \bra{0,N} \eta^{\dagger} \eta \ket{0,N}$, see Eqs.\ (\ref{eq:eta}) and (\ref{eq:alpha}), and can be expressed in terms of the isospin as follows:
\be
  \alpha 
  = 
  \frac1L
  \bra{0,N} \lr{\ff T^{2} - T_{z} (T_{z}+1)} \ket{0,N}
  \: .
\label{eq:alphaiso}  
\ee
Off half-filling, it is instructive to decompose the Hamiltonian $\ca H \equiv H - \mu \hat{N}$ into $\ca H = \ca H_{0} - (U-2\mu) T_{z} + \mbox{const.}$ where $\ca H_{0} = H - (U/2) \hat{N}$ is the charge-SU(2) symmetric part ($[\ff T, \ca H_{0}] = 0$) and where the second term represents a homogeneous field coupling to the $z$-component of $\ff T$.
The field explicitly breaks the charge-SU(2) symmetry and isospin-polarizes the ground state. 
We still have $[\ff T^{2} , \ca H] = 0$ and $[T_{z},\ca H] = 0$ (but $[\eta,\ca H] \ne 0$), i.e., like in a ferromagnet there are no fluctuations of $T_{z}$ and the ground state is characterized by the isospin quantum numbers $(T,M_{T})$ corresponding to $(\ff T^{2}, T_{z})$.
For fillings {\em above} half-filling, we have $\mu > U/2$, and the field term leads to a state with the isospin pointing into $-z$-direction, i.e., $M_{T} = - T$.
Using this in Eq.\ (\ref{eq:alphaiso}), we find
\be
  \alpha 
  = 
  - 2 M_{T} / L 
  = (N-L) / L
  = n-1
  \: , 
\label{eq:alphafin}  
\ee
where we also used $M_{T} = (L-N)/2$.
We conclude that the weight of the mode linearly increases from zero at half-filling to unity at $n=2$.
This explains the strong difference in the weights seen in the DMRG results (Figs.\ \ref{fig:AESk_n=1.8} and \ref{fig:AESk_n=1.1}).

The same analysis can be done for the $k$-resolved {\em two-particle} spectrum and fillings $n>1$.
Here, however, the exact eigenstate $\eta^{\dagger} \ket{0,N}$ would lead to a $\delta$-peak at {\em negative} frequency $\omega=\omega_{0}(\pi) = U-2\mu < 0$ for $n>1$ [which would contradict Eqs.\ (\ref{eq:A2particle}) and (\ref{eq:mu})] but the weight $\alpha = L^{-1} \langle \eta \eta^\dagger \rangle = L^{-1} \langle \lr{\ff T^2 - T_{z}(T_{z}-1)}\rangle$ vanishes exactly since $M_{T}=-T$.
For fillings $n<1$, the $\delta$-peak in the two-particle spectrum shows up on the $\omega>0$ side with a non-zero weight since $M_{T}=+T$ in this case -- analogous to the two-hole spectrum for $n>1$ and related to the latter by particle-hole transformation. 

Physically, the collective mode describes the {\em coherent} propagation of a doublon with energy $U-2\mu$ through the lattice without any scattering, i.e., the final state with wave vector $k=\pi$ in the two-hole spectroscopy has infinite lifetime. 
This is enforced by the remaining isospin-rotation symmetry around the $z$-axis which protects the ``$\eta$-mode'' at the zone edge from decay.
Note that the $\eta$-mode only shows up off half-filling and is not bound to off-diagonal long-range order as has been discussed previously. \cite{Yan89,Zha90}
For bipartite lattices of higher dimensions, the $\eta$-mode is found at the Brillouin-zone edges $\ff k=(\pm \pi, \pm \pi, ...)$. 

\section{Discussion}
\label{sec:con}

While the spectrum of one-particle excitations of the Hubbard model has been discussed extensively over the recent decades, much less is known for excitations involving two or more particles or holes.
Referring to the Lehmann representation of the spectral densities, it is obvious that the different spectroscopies are intimately related since they involve the same set of final states (when starting from a ground state with properly chosen total particle number). 
The main difference, however, is in the spectral weight attached to the final states. 
This strongly depends on the type of excitations studied.
Different spectroscopies therefore provide largely different perspectives on the same physics.

For strong $U$, the one-hole spectrum of the Hubbard model (corresponding to photoemission) exhibits the famous Hubbard bands. 
In the high-density regime (filling close to $n=2$), the lower Hubbard band is built from final states involving configurations with two holes at the same site. 
It is therefore shifted by about $U$ to higher binding energies and provides an interpretation of the Ni-6eV satellite, \cite{GBP+77} for example.
The weight of the satellite is small since the probability to excite the photoelectron from a singly occupied site is low for $n\to 2$. 

The same final states have a much higher weight in the two-hole spectrum (corresponding to Auger-electron spectroscopy) and the related peak is referred to as the Cini-Sawatzky satellite. 
This satellite in the two-hole spectrum can even be studied in the limit of the completely filled band ($n=2$) where, in the one-hole spectroscopy, the LHB disappears or has zero weight.
The related two-hole problem has attracted a lot of interest as it can be solved analytically and provides the concept of a repulsively bound pair of holes (called ``doublon'' in this paper).

Besides recent experiments addressing doublon physics in systems of ultracold atoms trapped in optical lattices, the two-hole problem has a longer history and has in particular stimulated much work on high-resolution Auger spectroscopy from metals. 
For $n<2$, however, a comprehensive overview of the two-hole excitations, or even an understanding of the spectral features has not yet been achieved. 
The obvious reason is that for $n<2$, one is confronted with an intractable many-body problem in general. 

For moderately large one-dimensional lattices, however, one can profit from the recent advances of numerical techniques based on matrix-product states. 
As is demonstrated here, this allows to study the two-hole excitation spectrum to a high degree of accuracy. 
We have employed a matrix-product state implementation of the density-matrix renormalization group in combination with the Chebyshev polynomial expansion technique to compute different excitation spectra. 
As the excitation is local ($k$-resolved spectra are computed by combining different local excitation spectra) the growth of the entanglement entropy with expansion order scales logarithmically and is very moderate (though it is larger for higher-order spectroscopies and quite substantial for a nearest-neighbor interaction $V \neq 0$).
Going to high orders (typically $M \sim 10^{3}$) provides us with the numerically exact solution on a frequency scale larger than a ``resolution'' $\delta E$ (typically $\delta E \sim 10^{-1}t$).

Computing the two-hole spectrum of the Hubbard model with this technique uncovers a couple of new phenomena. 
Some are intrinsically related to the one-dimensionality of the lattice, such as spin-charge separation.
These will be discussed in an upcoming paper but have been disregarded here as far as possible. 
With this we concentrate on those aspects which should, at least qualitatively, be the same for higher dimensions as well -- with the notable exception of physics related to ground states with spontaneously broken symmetries.

We find that the overall two-hole spectrum is governed by the physics of multiplons, i.e., bound states of infinite lifetime or resonances of finite spectral width showing up at characteristic excitation energies and with characteristic dispersions and weights.  
The footprints of multiplons are actually ubiquitous in the one-, two- and more-particle or hole spectra of the (extended) Hubbard model.
This multiplon physics is clearly developed at strong $U$ ($U = 6t = 1.5 W$ has been chosen throughout the paper) where the typical energy scales are well separated. 
In this regime some charge fluctuations can be disregarded, and effective models, like an effective hard-core-boson model (at high fillings) or the $t$-$J$ model (fillings close to half-filling), are very useful.

For $n=2$ and strong $U$, most of the weight of the two-hole spectrum is concentrated in the doublon peak. 
The two additional holes in the final state form a compound object of infinite lifetime at binding energies approximately centered around $U$ below the barycenter of the two-hole continuum.
For fillings $n\to 1$, however, the interpretation of the doublon peak as a repulsively bound hole pair changes, and it rather reflects the motion of an empty site through a lattice of singly occupied sites with antiferromagnetically correlated magnetic moments, as described by the $t$-$J$ model.

While the doublon is well defined in the whole filling range at $U=6t$, it cannot be expected to be stable (for $n < 2$) and should decay by emitting particle-hole excitations.
With decreasing $n$, the phase space for this decay channel increases; this destabilizes the doublon.
With increasing $U$, energy conservation requires the emission of multiple particle-hole excitations in a high-order scattering event; this stabilizes the doublon.
Besides these overall trends, however, the doublon lifetime has turned out to be strongly $k$-dependent. 
Interestingly, it is even infinite at $k=\pm \pi$ -- for any filling $n>1$ and for bipartite lattices of arbitrary dimension. 
This phenomenon is related to the ``hidden'' charge-SU(2) symmetry of the Hubbard model at half-filling. 
Off half-filling, the coupling of the chemical potential $\mu$ to the $z$-component of the isospin explicitly breaks this symmetry resulting in a massive collective mode in the two-hole spectrum (the doublon at $k=\pm \pi$) with energy $\omega=U-2\mu$ and with weight $\alpha=n-1$.

The two-hole spectrum also uncovers heavier multiplons which show up in certain filling ranges with lower weight.
Besides the doublon, we have identified a triplon and a quadruplon excitation. 
The triplon consists of a hole pair (a doublon) with an additional hole on a nearest-neighboring site.
The quadruplon is a pair of neighboring doublons. 
Let us note in passing that those can even be found, with extremely low weight at $U=6t$, in the one-hole spectrum. 
Furthermore, the multiplon hierarchy can also be continued:
There is for example evidence for a quintuplon, i.e., five holes on three sites, showing up as a weak peak on the high-binding-energy side of the quadruplon in the four-hole spectrum (see Appendix \ref{sec:four}).

For a Hubbard-type model with local interaction and with two (spin) degrees of freedom per site, an odd-even effect is a natural expectation. 
It would seem that multiplons composed of an odd number of holes do neither form stable compounds nor well-defined resonances unless a nearest-neighbor density interaction $V$ is added to the Hamiltonian.
Contrary, the doublon and the quadruplon appear as resonances with comparably small width or even as stable objects with infinite lifetime in the $k$-resolved two-hole spectrum.
The four-hole spectrum at $n=2$ represents an interesting special case, where a stable quadruplon is found in a finite $k$-range around $k=\pm \pi$ until it is immersed in the doublon-doublon continuum and becomes a sharp resonance of finite width.
Odd multiplons, when stabilized by $V$, possess one or more internal degrees of freedom as there are hopping processes which do not affect their integrity in case of strong interactions.
In the case of the triplon, for example, this manifests itself in a double-peak substructure on a scale $2t$. 

It is important to distinguish clearly between dispersion and decay. 
The width of multiplon peaks in the local two-hole spectrum is mainly due to dispersion, which is finite for finite $U<\infty$ and can be understood best in the framework of effective models obtained from strong-coupling perturbation theory. 
Also, $k$-resolution helps to identify the intrinsic lifetime broadening of resonances.

To distinguish between resonances and continua can be problematic in some cases.
Here, the triplon and the quadruplon represent good examples, as their excitation energies lie within the doublon-hole or within the doublon-doublon continuum, respectively.
However, for $V=0$ only the quadruplon is a compound object with finite lifetime, while the triplon decays immediately.
To analyze such cases and come to unambiguous conclusions, we have applied a new filter-operator technique.
This allows to extract information on the states contributing to the spectrum at a given frequency, which is specifically dependent on the choice of an arbitrary filter operator.
Within the framework of Chebyshev polynomial expansion, this can be implemented in a straightforward way.

There are several lines along which one can extend the present study, as for example study the manifestation of spin-charge separation in the two-hole spectrum, or the footprints of spontaneous U(1)-symmetry breaking for $U<0$. 
Another set of questions regards the real-time dynamics of multiplons and the relation to time-dependent (pump-probe) Auger-electron spectroscopy. 
Work in these directions is in progress.

\acknowledgments

Support of this work by the Deutsche Forschungsgemeinschaft within the SFB 925 (project B5) is gratefully acknowledged.

\appendix

\section{(Inverse) photoemission spectra} 
\label{sec:pes}

\begin{figure}[b]
\includegraphics[width=\columnwidth]{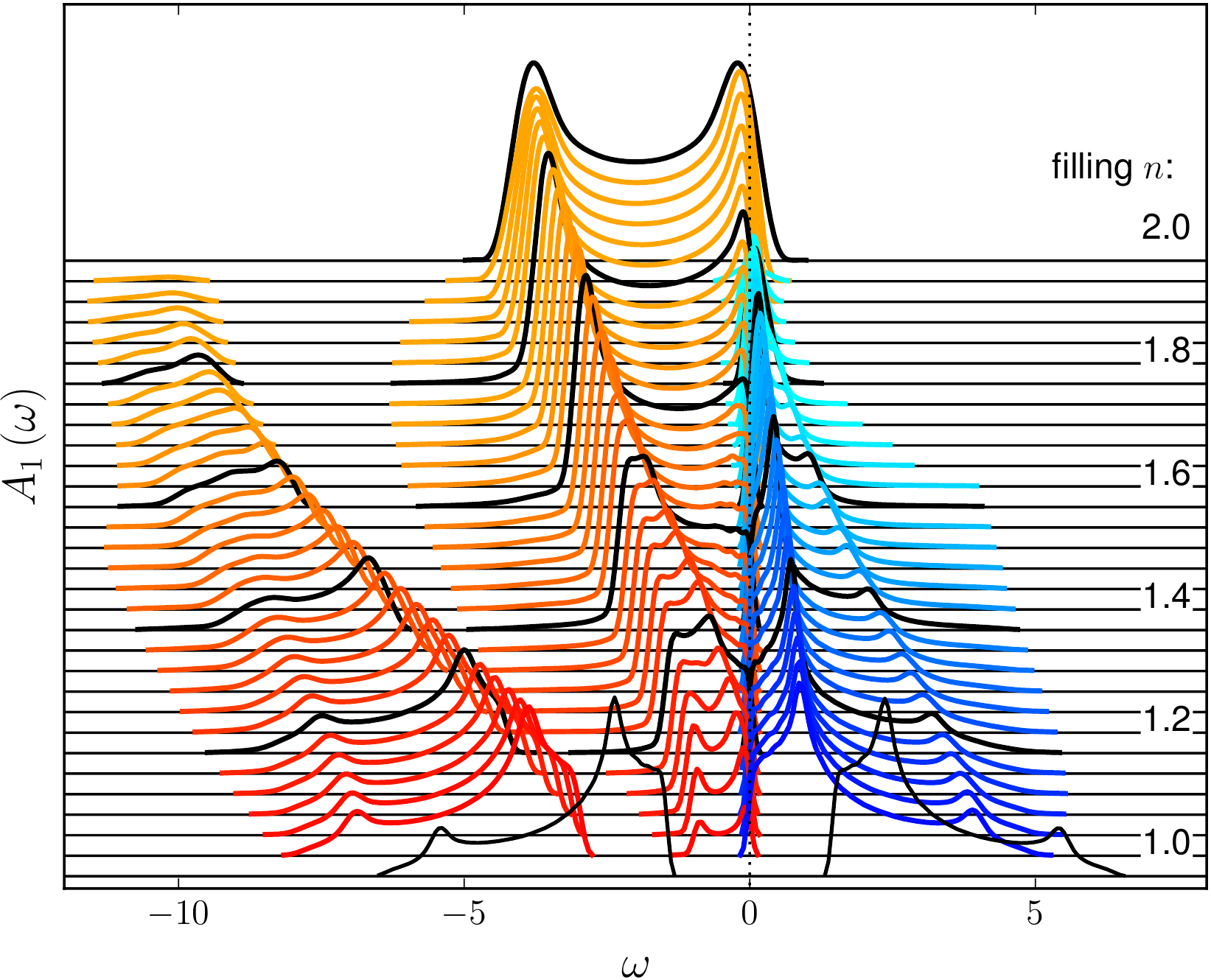}
\caption{
One-hole excitation spectrum [Eq.\ (\ref{eq:A1hole})] for $\omega<0$ and one-particle spectrum for $\omega>0$ [Eq.\ (\ref{eq:oneparticle})] for $U=6t$ and various fillings $n$. 
The filling values indicated on the right are drawn in black.
$\delta E=0.2t$ (up to $M=1154$ Chebyshev moments).
Other parameters as in Fig.\ \ref{fig:nseries}. 
}
\label{fig:pes}
\end{figure}

Fig.\ \ref{fig:pes} shows the (local) one-hole excitation spectrum [see Eq.\ (\ref{eq:A1hole})] and the one-particle excitation spectrum, 
\begin{eqnarray}
A_{{\rm 1-particle},\sigma}\lr{\omega} 
&=& 
\sum_n \norm{\matrixel{n,N+1}{c^{\dagger}_{i\sigma}}{0,N}} 
\nonumber \\
&\times&
\delta\lr{\omega + \mu - (E_n^{\lr{N+1}}-E_0^{\lr{N}})}
\: . 
\nonumber \\
\label{eq:oneparticle}
\end{eqnarray}
at $U=6t$ and for fillings ranging from half-filling ($n=1$) up to limit of the completely filled band ($n=2$). 
These correspond to photoemission (one-hole spectrum) and inverse photoemission (one-particle spectrum).

For $n=2$, the unrenormalized density of states is recovered: $A_{{\rm 1-particle},\sigma}\lr{\omega} = \rho_{0}(\omega + \mu)$. 
The van Hove singularities at $\omega=0$ and $\omega=-4t$ are somewhat broadened due to the finite $\delta E = 0.2t$. 
For fillings $n<2$, the spectrum consists of the upper Hubbard band (UHB) at low binding energies and the lower Hubbard band (LHB) showing up at higher binding energies, approximately at $U$ below the UHB.
Note the strong filling-dependent spectral-weight transfer. 
At half-filling the LHB and the UHB are placed symmetrically to $\omega=0$.
The substructure of the LHB and UHB can be analyzed and understood in terms of collective spinon and holon excitations, see Refs.\ \onlinecite{BGJ04,FH09,Koh10}.

\section{Four-hole spectra} 
\label{sec:four}

\begin{figure}[b]
\includegraphics[width=\columnwidth]{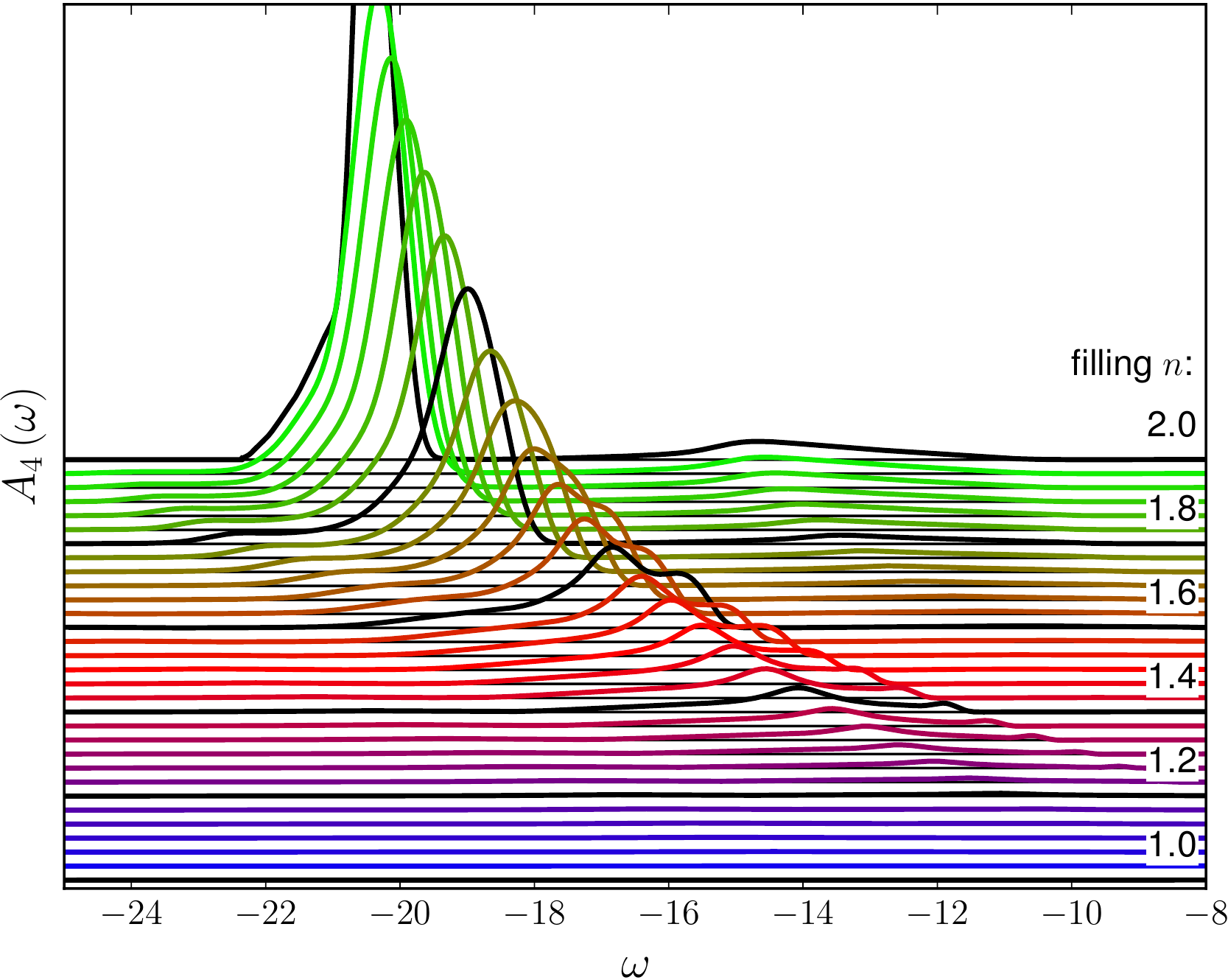}
\caption{
Four-hole excitation spectrum [Eq.\ (\ref{eq:fourhole})] for $U=6t$ and various fillings $n$.
The filling values indicated on the right are drawn in black.
Other parameters as in Fig.\ \ref{fig:nseries}.
}
\label{fig:four}
\end{figure}

Fig.\ \ref{fig:four} shows the (local) four-hole excitation spectrum at $U=6t$ for various fillings in the range of excitation energies $\omega$ where the structure ``Q'' is seen in the two-hole spectrum (Fig.\ \ref{fig:nseries}). 
The position of the main peak in Fig.\ \ref{fig:four} approximately coincides with the position of ``Q'' in the two-hole spectrum for all fillings $n$, if the same final subspaces are compared (for example, at $\omega \approx -14.5t$ in $A_{\rm 2-hole}\lr{\omega}$ for $n=84/60=1.4$ compared to $\omega \approx -14.6t$ in $A_{\rm 4-hole}\lr{\omega}$ for $n=86/60\approx 1.43$).
For $n\to 2$, the main peak evolves into the quadruplon excitation that has been analyzed in Sec.\ \ref{sec:quad}. 
The weight of the quadruplon strongly depends on the spectroscopy considered. 
In the two-hole spectrum, it is most intense for $n \to 1$, while it has maximum weight for $n\to 2$ in case of the four-hole spectroscopy.

The latter also uncovers the existence of a ``quintuplon'', i.e., a final state composed of two neighboring doublons plus an additional neighboring hole. 
This is visible as the additional structure on the high-binding-energy side of the quadruplon in Fig.\ \ref{fig:four} (see $\omega \approx - 22t$ for $n=1.8$, for example). 
As in the triplon case, we suspect that the quintuplon can be stabilized as a truly bound state with infinite lifetime when switching on a nearest-neighbor Coulomb interaction. 
For $V=0$, however, it is not stable and should correspond to a hole-quadruplon continuum of final states.


\begin{thebibliography}{63}
\bibitem[{mu()}]{mu}
\bibinfo{note}{For a finite system, $\mu$ not unambiguously defined. With Eq.\
  (\ref{eq:mu}) we here used here a symmetric definition, where the chemical
  potential for $N$ particles is given as a difference of the total energies
  for $N\pm 1$ particles. This ensures that the excitation energies in the
  two-hole (two-particle) spectrum are strictly non-positive (non-negative).
  The choice implies that the two branches of the spectrum are computed with
  (slightly) different chemical potentials. Hence, it cannot be used for $n=1$
  where the function $\mu=\mu(n)$ is discontinuous.}
\end{thebibliography}
\end{document}